\documentclass[10pt, preprint]{aastex}

\newcommand{\dsf}[2]{\displaystyle{\frac{#1}{#2}}}
\slugcomment{\scriptsize{REVISED VERSION DECEMBER 12, 2005.
Preprint typeset using \LaTeX style}}
\shorttitle{MHD INSTABILITY IN PROTO-NEUTRON STARS}
\shortauthors{Masada et al.}
\begin{document}
\title{Nonaxisymmetric Magnetorotational Instability in Proto-Neutron
  Stars}

\author{Youhei Masada\altaffilmark{1,2}, Takayoshi Sano\altaffilmark{3},
  and Hideaki Takabe\altaffilmark{3}}

\altaffiltext{1}{Kwasan and Hida Observatories, Kyoto University, Yamashina, 
Kyoto 607-8471, Japan; masada@kusastro.kyoto-u.ac.jp}
\altaffiltext{2}{Department of Astronomy, Kyoto University, Sakyo, Kyoto 606-8502, Japan}
\altaffiltext{3}{Institute of Laser Engineering, Osaka University, Suita, 
Osaka 565-0871, Japan} 
\begin{abstract}
We investigate the stability of differentially rotating proto-neutron
stars (PNSs) with a toroidal magnetic field.
Stability criteria for nonaxisymmetric MHD instabilities are
derived using a local linear analysis.
PNSs are expected to have much stronger radial shear in the rotation
velocity compared to normal stars.
We find that nonaxisymmetric magnetorotational instability (NMRI)
with a large azimuthal wavenumber $m$ is dominant over the kink mode
($m=1$) in differentially rotating PNSs.
The growth rate of the NMRI is of the order of the angular velocity
$\Omega$ which is faster than that of the kink-type instability 
by several orders of magnitude.
The stability criteria are analogous to those of the axisymmetric
magnetorotational instability with a poloidal field, although the
effects of leptonic gradients are considered in our analysis.
The NMRI can grow even in convectively stable layers if the
wavevectors of unstable modes are parallel to the restoring force
by the Brunt-V\"ais\"al\"a oscillation.
The nonlinear evolution of NMRI could amplify the magnetic fields and
drive MHD turbulence in PNSs, which may lead to enhancement of the 
neutrino luminosity.
\end{abstract}

\section{INTRODUCTION}
Neutron stars are formed in the aftermath of supernova explosions
associated to the gravitational core-collapse of 
massive stars (8 -- $30M_{\sun}$) at the end of their evolutions. 
The binding energies released by core-collapse are stored in the
interior of a newly born hot neutron star, and those energies are
emitted as the neutrinos in the explosion stage. 
Neutrino emission from proto-neutron stars (PNSs) is one of the key
processes of the delayed explosion scenario of core-collapse
supernovae, in other words, the size of neutrino luminosities and the  
energy deposition efficiency determine whether the explosion will
succeed or not (Bethe 1990). 

Convection in PNSs is considered to play 
an important role in enhancing the neutrino luminosities 
(Epstein 1979; Lattimer \& Mazurek 1981; Arnett 1987; 
Burrows \& Lattimer 1986; 
Burrow \& Fryxell 1993). 
Other hydrodynamic instabilities may also contribute to 
the enhancement of neutrino luminosities 
(Bruenn \& Dineva 1996; Mezzaacappa et al. 1998a; 
Miralles et al. 2000, 2002, 2004). 
Mayle (1985) and Wilson \& Mayle (1988, 1993) have argued that 
the material in the PNSs would be unstable to a double diffusive 
instability which is referred to as ``neutron finger''. 
Bruenn et al. (2004) have found a new double diffusive 
instability called ``lepto-entropy finger instability''. 
However, despite considering various hydrodynamic instabilities, 
recent numerical simulations have not succeeded in producing
sufficient neutrino luminosities leading to the delayed 
supernova explosions (Burrows et al. 1993; Keil et al. 1996; 
Mezzacappa et al. 1998b; Buras et al. 2003). 

One of the primary causes of the inefficient neutrino luminosity 
is the strong radial gradients of the entropy and leptonic fraction.
Stable stratification of these quantities appears at the outer layer
of PNSs and suppresses convective motions (Janka \& M\"uller 1996). 
Convective regions are located deep inside of PNSs
and surrounded by a convectively stable layer 
in which the surface fluxes of $\nu_e$ and $\tilde{\nu_e}$ 
are mainly built up (Buras et al. 2003). 
Thus the convection in PNS has little influence on the emission of
neutrinos and is irrelevant for the supernova dynamics.

On the other hand, MHD instabilities could be a seed 
of turbulence even in the stably stratified layer. 
Magnetic effects on the dynamics of supernovae have been 
examined actively since the importance of magnetorotational
instability (MRI) in supernova cores is pointed out (Akiyama et
al. 2003).
Two-dimensional MHD simulations of 
the collapse of supernova cores indicate that the shapes of
shock waves and the neutrino sphere can be modified by the effects of
magnetic fields (Kotake et al. 2004; Yamada \& Sawai 2004; Takiwaki et
al. 2004). 
Thompson et al. (2005) suggest that 
the turbulent viscosities sustained by the MRI convert 
the rotational energies of supernova core into the thermal energies 
and may affect the supernova dynamics. 
However, effects of magnetic phenomena inside of the PNSs
have not been understood yet and still have many uncertainties. 

Therefore, in this paper we analyze the linear growth of 
MHD instabilities in the stably stratified layer of PNSs, 
and discuss the possibility of enhancement of neutrino luminosities.
Especially, we focus on the instabilities caused by a strong
differential rotation. 
We assume the toroidal component of the field is dominant in the
supernova cores. 
Local stability of stellar objects with a rotation and toroidal 
magnetic fields have been investigated by many authors
(Fricke 1969; Tayler 1973; Acheson 1978; Pitts \& Tayler 1985; 
Schmitt \& Rosner 1986; Spruit 1999, 2002). 
Spruit (1999) concluded that the kink-type instability (Tayler
instability) grows at first among various MHD instabilities in stellar
radiative zone. 
However, most of the previous work assume a weak differential rotation
because their motivation is for the interiors of main sequence stars
like the Sun.
The stability analysis for the cases of a strong differential
rotation, which is expected in PNSs, has not been carried out
sufficiently. 
In such cases, nonaxisymmetric magnetorotational instability (NMRI)
could be dominant over the kink-type instability (Balbus \& Hawley 1992). 
The nonlinear evolutions of the NMRI can initiate and sustain MHD
turbulence and amplify magnetic fields, 
which may lead to the enhancement of neutrino luminosity 
via magnetoconvection, viscous heating, and the other MHD phenomena. 

The paper is organized as follows. 
In \S2 we obtain the dispersion equation
that determines the stability of PNSs. 
In \S3 we discuss the general properties of the dispersion relation, 
and in \S4 stability criteria and the maximum growth rate are derived 
by an analytical approach. 
In \S5 we apply our results to the interiors of PNSs, 
and examine the linear growth of MHD instabilities. 
The conditions that the NMRI can be dominant over 
the kink-type instability and 
the nonlinear effects of MHD instabilities are discussed in \S6. 
Finally, we summarize our main findings in \S7.

\section{THE DISPERSION RELATION}
We examine the stability of PNSs with a toroidal magnetic field using 
a linear perturbation theory. 
Acheson (1978) obtained a local dispersion equation for MHD 
instabilities in stellar objects with a toroidal field. 
We derive our dispersion equation following Acheson's analysis. 
One of the original features of this work is that we consider the effects 
of leptonic gradients which can be important in PNSs. 
The size of differential rotation in PNSs could be much larger 
than that in other stellar objects (Villain et al. 2004). 
Therefore we focus on the cases of a strong differential rotation which 
are not well studied in previous work.

The leptonization and cooling timescales of PNSs are much longer than 
the growth time of the instabilities considered in this paper. 
Thus the unperturbed state can be treated in a 
quasi-stationary approximation. 
The conductivity of plasma inside PNSs is so large that the decay 
timescale of magnetic fields is much longer than their cooling time. 
Then, the governing equations are the ideal MHD equations; 
\begin{equation}
\displaystyle{\frac{\partial\rho}{\partial t}} 
 + \nabla \cdot (\rho \mbox{\boldmath $u$}) = 0 \;, 
\label{eq1}
\end{equation}
\begin{equation}
\displaystyle{\frac{\partial \mbox{\boldmath $u$}}{\partial t}} 
+ ( \mbox{\boldmath $u$}\cdot
\nabla ) \mbox{\boldmath $u$} = - \frac{1}{\rho }\nabla
\left( p+\frac{1}{2\mu }\mbox{\boldmath $B$}^2 \right) 
+ \frac{1}{\rho}( \mbox{\boldmath $B$}\cdot\nabla ) 
\mbox{\boldmath $B$} + \mbox{\boldmath $g$} \;,
\label{eq2} 
\end{equation}
\begin{equation}
\displaystyle{\frac{\partial\mbox{\boldmath $B$}}{\partial t}} = 
\nabla \times (\mbox{\boldmath $u$}\times\mbox{\boldmath $B$}) \;, 
\label{eq3} 
\end{equation}
where $\rho$ is the mass density, 
$\mbox{\boldmath $u$}$ is the fluid velocity, $p$ is the pressure, 
$\mbox{\boldmath $B$}$ is the magnetic field, 
$\mbox{\boldmath $g$}$ is the gravity, and 
$\mu$ is the magnetic permeability.

We use the cylindrical coordinates $(r, \phi, z)$, 
and consider the Eulerian perturbations (denoted 
by a prefix $\delta $) with the WKB spatial and temporal dependence, 
$\delta \propto \exp \{i (k_r r + m\phi + k_z z - \sigma t)\}$. 
Our analysis is purely local one, formally valid only in the 
immediate neighborhood of a particular location $(r,z)$ and then only 
for perturbations of wavelength $\lambda$ in the $r$-$z$ plane such that 
$\lambda^2 \ll r^2 + z^2 $. 
The azimuthal wavelength can be quite large, perhaps order of
radius. 

Let us consider a PNS rotating with the angular velocity $\Omega$ and
toroidal magnetic fields $B_{\phi}$. 
We assume these variables depend only on $r$ and have a power-law 
dependence; 
$\Omega (r) \propto r^{-q}$ and $B_{\phi} (r) \propto r^{-s}$. 
At the unperturbed state, the PNS is in the magneto-hydrostatic 
equilibrium, that is, 
\begin{equation}
\displaystyle{\frac{B_{\phi }^2}{\mu\rho r}} +
\displaystyle{\frac{1}{\rho}}\displaystyle{\frac{\partial p}{\partial r}}+
\displaystyle{\frac{B_{\phi }}{\mu \rho}}
\displaystyle{\frac{\partial B_{\phi}}{\partial r}} = 
-g_r + r\Omega^2 \;,
\label{eq4} 
\end{equation}
\begin{equation}
\vbox{\kern 20pt} \displaystyle{\frac{1}{\rho}}
\displaystyle{\frac{\partial p}{\partial z}} = -g_z \;,
\label{eq5}
\end{equation}
in the radial and vertical directions, respectively. 

When written out in component form and only the largest terms 
retained, the above set of equations becomes to linear order ;
\begin{equation}
k_r\delta u_r + k_z\delta u_z  =  0 \;, 
\label{eq6} 
\end{equation}
\begin{equation}
i\omega\delta u_r + 2\Omega\delta u_{\phi} +
\displaystyle{\frac{imB_{\phi}}{\mu\rho r}}\delta b_r = 
\displaystyle{\frac{ik_r}{\rho}}\delta p +
\displaystyle{\frac{B_{\phi}}{\mu\rho r}}(2+ik_r r)\delta b_{\phi} +
\displaystyle{\frac{\delta\rho}{\rho}}g_r^* \;, 
\label{eq7} 
\end{equation}
\begin{equation}
i\omega\delta u_{\phi} - (2 - q)\Omega\delta u_r =
 - (1 - s)\displaystyle{\frac{B_{\phi}}{\mu\rho r}}\delta b_r +
\displaystyle{\frac{im}{\rho r}}\delta p \;, 
\label{eq8} 
\end{equation}
\begin{equation}
i\omega\delta u_z + \displaystyle{\frac{imB_{\phi}}{\mu\rho r}}\delta
B_z = \displaystyle{\frac{ik_z}{\rho }}\delta p +
\displaystyle{\frac{ik_z B_{\phi}}{\mu\rho}}\delta b_{\phi} +
\displaystyle{\frac{\delta\rho}{\rho }}g_z^* \;, 
\label{eq9} 
\end{equation}
\begin{equation}
\omega\delta b_r = -\displaystyle{\frac{mB_{\phi}}{r}}\delta u_r \;,
\label{eq10} 
\end{equation}
\begin{equation}
i\omega\delta b_{\phi} + (1 + s
)\displaystyle{\frac{B_{\phi}}{r}}\delta u_r = q\Omega\delta b_r -
\displaystyle{\frac{imB_{\phi}}{r}}\delta u_{\phi } \;, 
\label{eq11} 
\end{equation}
\begin{equation}
\omega\delta b_z = -\displaystyle{\frac{mB_{\phi }}{r}}\delta u_z \;,
\label{eq12}
\end{equation}
where  
$$
\omega = \sigma - m\Omega \;, 
$$
$$
\mbox{\boldmath $g$}^* = (- g_r^*, - g_z^*) = (- g_r + r\Omega^2,
-g_z) \;. 
$$
Here we assume the local approximation $(k \gg 1/r, m/r)$. 
We also adopt the Boussinesq approximation (e.g.,
Chandrasekhar 1961) because the Alf\'ven speed and the rotation 
velocity of PNSs would be much smaller than the sound speed $c_s
\simeq 0.1c$ where $c$ is the speed of light.
In addition to these linearized equations, 
we use a relation by virtue of the local approximation, 
$\delta p + B_{\phi}\delta b_{\phi}/\mu  = 0$. 
This relation means that a fluctuation of the 
total pressure is negligible (Acheson 1978). 

To complete our set, we require another linearized equation. 
The chemical equilibrium is realized in PNSs, and thus the density is 
given by a function of pressure $p$, 
temperature $T$, and leptonic fraction $Y_l = (n_e + n_{\nu})/n$ where
$n_e$, $n_{\nu}$, and $n$ are the number densities of 
electrons, neutrinos, and baryons, respectively.
In the Boussinesq approximation, 
the pressure perturbations are
negligible because fluid elements are assumed to be moving slowly and 
in the pressure equilibrium with their surroundings. 
Then, the perturbations of density $\delta \rho$ and entropy per
baryon $\delta S$ can be expressed in terms of the perturbations of
temperature $\delta T$ and leptonic fraction $\delta Y_l$: 
\begin{equation}
\delta \rho =  - \rho\Big(\alpha \displaystyle{\frac{\delta T}{T} 
+ \xi\delta Y_l}\Big) \;, 
\label{eq15} 
\end{equation}
\begin{equation}
\delta S = m_B c_p \dsf{\delta T}{T} + \zeta \delta Y_l \;, 
\label{eq15a}
\end{equation}
where $m_B$ is the baryon mass, $\alpha=-(\partial\ln\rho /\partial\ln
 T)_{pY_l}$, $\xi = -(\partial\ln\rho/\partial Y_l)_{pT}$
are the coefficients of thermal and chemical expansion, and 
$c_p = (T/m_B)(\partial S/\partial T)_{pY_l}$ is the specific heat at
 constant pressure. 
The sign of $\zeta =(\partial S/\partial Y_l)_{pT}$ 
determines whether the leptonic gradient is stabilizing or destabilizing. 
Here we neglect the effects of thermal and chemical diffusions for 
simplicity. The role of these diffusive processes on the growth of MHD 
instabilities is discussed later in \S~\ref{sec:discussion}.1. 
The equations governing the thermal balance and the diffusion of 
leptonic fraction are written as
\begin{equation}
\dsf{\partial S}{\partial t} + \mbox{\boldmath $u$} \cdot \nabla S = 0
\;, 
\label{eq16d} 
\end{equation}
\begin{equation}
\dsf{\partial Y_l}{\partial t} + \mbox{\boldmath $u$} \cdot \nabla Y_l
= 0 \;, 
\label{eq17d} 
\end{equation}
(Miralles et al. 2004) and then the linearized equations are given by
\begin{equation}
-i\omega\delta T - \delta \mbox{\boldmath $u$} \cdot
 \left[\left(\dsf{\partial T}{\partial p} \right)_{s,Y_l} \nabla p - \nabla T
 \right] = 0 
 \;, 
\label{eq16} 
\end{equation}
\begin{equation}
-i\omega\delta Y_l + \delta \mbox{\boldmath $u$} \cdot \nabla Y_l =
 0 \;. 
\label{eq17}
\end{equation}
Substituting equations (\ref{eq16}) and (\ref{eq17}) into equation
(\ref{eq15}), we obtain the linearized equation, 
\begin{equation}
i\omega\displaystyle{\frac{\delta\rho}{\rho}} + 
\delta\mbox{\boldmath $u$}\cdot \nabla \Psi = 0 \;, 
\label{eq18}
\end{equation}
where 
\begin{equation}
\nabla\Psi = - \dsf{\alpha }{T}\left[\Big(\displaystyle
{\frac{\partial T}{\partial p}}
\Big)_{s,Y_l}\nabla p - \nabla T \right] + \xi \nabla Y_l \;. 
\label{eq19}
\end{equation} 

Eliminating perturbed quantities in eight linearized equations
[eqs. (\ref{eq6}) -- (\ref{eq12}) and (\ref{eq18})], we obtain the
following fourth-order dispersion equation, 
\begin{eqnarray}
\displaystyle{\frac{k^2}{k^2_z}}\omega^4 - 
\left[ 2(1 + s)\omega^2_A +
  \kappa^2 + 2{\frac{k^2}{k^2_z}}m^2\omega^2_A + 
\dsf{k^2}{k^2_z}\mbox{\boldmath $G$}\cdot\nabla\Psi 
\right] \omega^2
- 8m\omega^2_A\Omega\omega  \nonumber \\
- m^2\omega^2_A \left[ 2q \Omega^2 + 2(1 -
  s)\omega^2_A - {\frac{k^2}{k^2_z}}m^2\omega^2_A 
- \dsf{k^2}{k^2_z}\mbox{\boldmath $G$}\cdot\nabla\Psi \right] = 0 \;, 
\label{eq20} 
\end{eqnarray}
where
\begin{equation}
k^2 = k^2_r + k^2_z \;, \nonumber 
\end{equation}
\begin{equation}
\kappa^2 = 2(2 - q)\Omega^2 =
\displaystyle{\frac{1}{r^3}}{\frac{\partial(r^2\Omega)^2}{\partial
    r}} \;, \nonumber 
\end{equation}
\begin{equation}
\omega_A =
\displaystyle{\frac{B_{\phi}}{\sqrt{\mu\rho}r}}  =
\displaystyle{\frac{v_{A\phi}}{r}} \;,  \nonumber 
\end{equation} 
and 
\begin{equation}
\mbox{\boldmath $G$} = \mbox{\boldmath $g$}^* - \mbox{\boldmath
  $k$}(\mbox{\boldmath $k$}\cdot\mbox{\boldmath $g$}^*)/k^2\;, 
\end{equation} 
are the square of a poloidal wavenumber, the square of the 
epicyclic frequency, the Alf\'en frequency, and the 
effective gravity vector. 
When we assume that isobaric and isochoric surfaces coincide, 
the dispersion equation (\ref{eq20}) is simplified further by noting that 
\begin{equation}
\mbox{\boldmath $g$}^* \times \nabla\Psi = g_r^*\nabla_z\Psi   
-  g_z^*\nabla_r\Psi = 0
\;. 
\label{eq21}
\end{equation}
Then, defining $N^2_r=g^*_r \nabla_r \Psi$
with an analogous expression for $N^2_z$,  
the final form of the dispersion equation becomes, 
\begin{eqnarray}
\displaystyle{\frac{k^2}{k^2_z}}\omega^4 - 
\left[2(1 + s)\omega^2_A + \kappa^2 + 2{\frac{k^2}{k^2_z}}m^2\omega^2_A 
+ N^2_e\right]
\omega^2 - 8m\omega^2_A\Omega\omega  \nonumber \\
- 2qm^2\omega^2_A\Omega^2 - 2 (1 - s)m^2\omega^4_A +
{\frac{k^2}{k^2_z}} m^4\omega^4_A + m^2\omega^2_A N^2_e =  0 \;,
\label{eq22}
\end{eqnarray}
where
$$
N^2_e = \left(N_r - \dsf{k_r}{k_z}N_z\right)^2 \;.
$$
The quantities $N_r$ and $N_z$ are components of 
the Brunt-V\"ais\"al\"a frequency 
$N = (\mbox{\boldmath $g$}^* \cdot \nabla \Psi)^{1/2}$. 
As the leptonic gradient is considered here, 
the Brunt-V\"ais\"al\"a frequency is divided into 
the thermal buoyancy frequency $N_T$ and the leptonic buoyancy
frequency $N_{Y_l}$;
\begin{equation}
N^2 = N^2_T + N^2_{Y_l}\;,
\end{equation}
where 
\begin{equation}
N^2_T = - \dsf{\alpha \mbox{\boldmath $g$}^*}{T}\cdot
\left[\left(\dsf{\partial T}{\partial p} \right)_{s,Y_l}\nabla p -
  \nabla T\right], \ \ \ \ \ 
N^2_{Y_l} = \mbox{\boldmath $g$}^* \cdot \xi \nabla Y_l \;.
\end{equation}

If the leptonic gradient is ignored, 
the dispersion equation (\ref{eq22}) is identical to that obtained by
Acheson (1978) in the incompressible limit ($c_s \to \infty $). 
There are two physical branches to the dispersion equation: 
An internal gravity wave branch which is present 
in the absence of magnetic fields, and 
an Alfv\'en wave branch which becomes unstable 
at sufficiently long azimuthal wavelengths 
when magnetic fields are present. 

Notice that the local analysis is irrelevant in the case 
that the growth rate of unstable modes is smaller than the
differential rotation rate $| d \Omega
/ d \ln r | = q \Omega$. 
In that case, this problem should be treated as an initial-value 
problem (Balbus \& Hawley 1992; Kim \& Ostriker 2000). 
This is because nonaxisymmetric disturbances cannot have a simple 
plane waveform in the presence of the shearing background on the wave 
crests (Goldreich \& Lynden-Bell 1965). 
In this paper, we focus mainly on the NMRI for strong shear cases ($q
\sim 1$) of which the growth rate is comparable to the angular
velocity and use the local dispersion equation (\ref{eq22}) for the
stability analysis for simplicity.
Kim \& Ostriker (2000) examine the linear growth of the NMRI using
both the local WKB analysis and direct integration of the linearized
shearing-sheet equations taking account of the time-dependence of the
radial wavenumber.
They found that the instantenious growth rate in the direct
integration is in good agreement with the WKB dispersion relation.
The advantage of the simple WKB analysis is that we can derive analytical
formula of the stability criteria and the growth rate of the
instabilities.
In fact, the stability conditions obtained by the WKB dispersion
equation for a case of weak differential rotation (Acheson 1978) are
quite useful and applied to the recent studies of the evolution of
massive stars (Spruit 1999; Heger et al. 2005).

\section{GENERAL FEATURES OF THE DISPERSION RELATION}
In this section we solve the dispersion equation derived in the
previous section and examine the characteristics of MHD instabilities
in PNSs.
There are four important parameters in the dispersion equation
(\ref{eq22}): the shear parameter $q$, the Brunt-V\"ais\"al\"a frequency
$N$, the ratio between the angular velocity and the Alfv\'en frequency
$\Omega /\omega_A$, and the power index of the field distribution $s$.
The nature of MHD instabilities, such as the growth rate and
characteristic wavelength, depends on these parameters, especially on
the shear parameter $q$ and the Brunt-V\"ais\"al\"a frequency $N$.

The shear parameter is quite small in the Sun where the kink
instability would be dominant among MHD instabilities (Spruit 1999).
However the shear parameter can be of the order of unity in PNSs.
In that case, the magnetorotational instability (MRI) can be important
as is in rotationally supported accretion disks (e.g., Balbus \& Hawley
1998).
The growth rate of the MRI could be faster than that of the kink
instability by many orders. 
MHD turbulence generated by the MRI can transport angular momentum
inside of PNSs, which may affect their dynamical evolutions.

In stably stratified regions, the Brunt-V\"ais\"al\"a frequency $N$
is real and denotes the size of the negative buoyancy. 
When $N$ is large, the growth of the axisymmetric MRI is suppressed
significantly because the Brunt-V\"ais\"al\"a oscillation prevents the
radial motion of fluids (Balbus \& Hawley 1994).
If the field is purely azimuthal, axisymmetiric modes are stable for
the MRI and only nonaxisymmetric modes can be unstable.
The nonaxisymmetric MRI (NMRI) in accretion disks has been examined
(Balbus \& Hawley 1992; Terquem \& Papaloizou 1996;
Kim \& Ostriker 2000) and the operating mechanism is described in
detail by Kim \& Ostriker (2000). 
In this paper we investigate the stabilizing effects of $N$ on the growth
of the NMRI.

In what follows, we fix the values of the other two parameters. 
The ratio of angular velocity to the Alfv\'en frequency and the field 
distribution are assumed to be $\Omega /\omega_A = 30$ and $s = -2$,
respectively.
These are about their typical values in rotating PNSs (see
\S~\ref{sec:application}). 

\subsection{Effects of the Shear Parameter}
\begin{figure}
\epsscale{.80}
\plotone{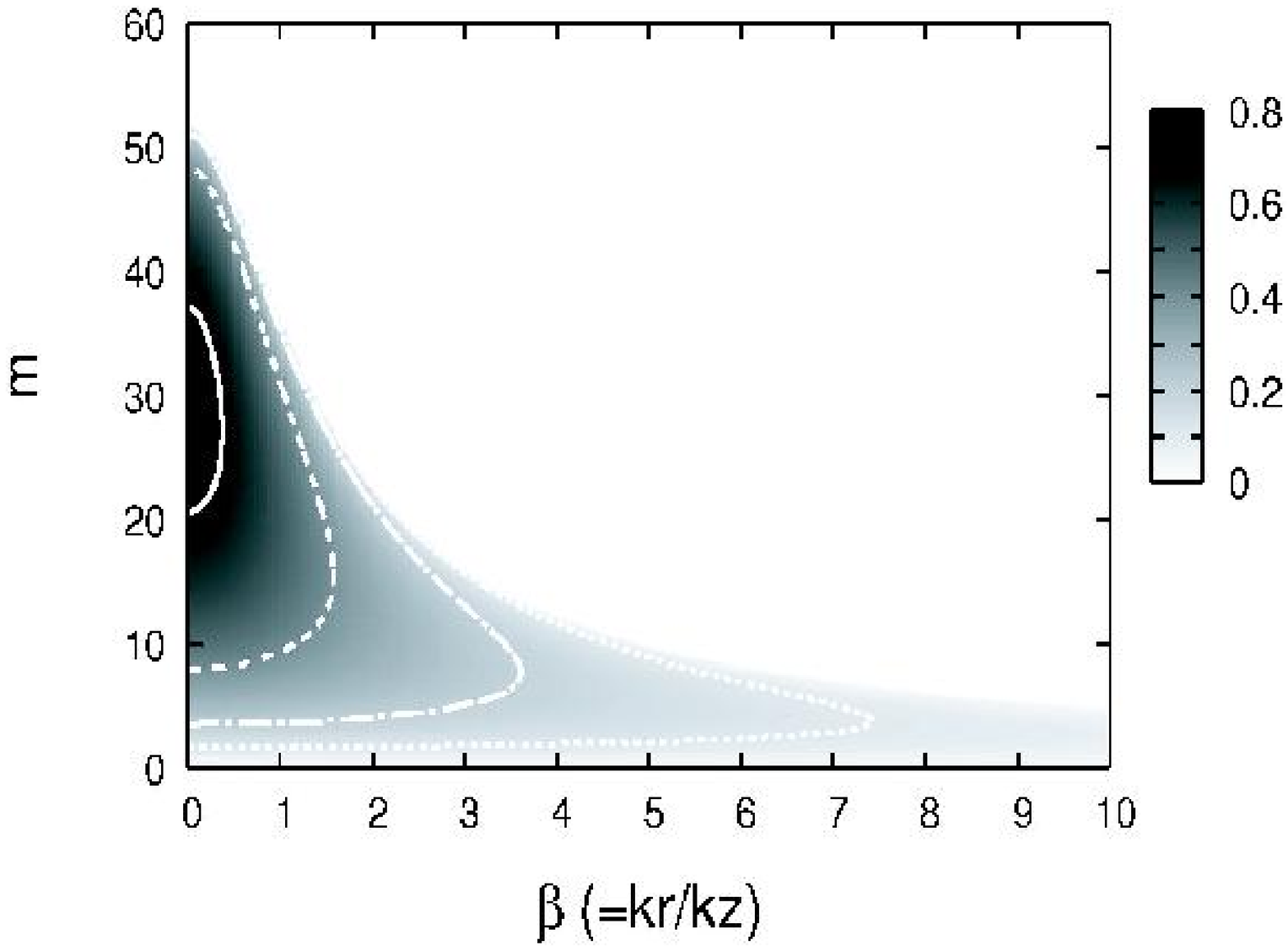}
\caption{The unstable growth rate $\gamma \equiv \Im (\omega)$ is plotted by
the gray contour on the $m$-$\beta$ plane, where $m$ is the azimuthal
wavenumber and $\beta \equiv k_r/k_z$ denotes the ratio of the
poloidal wavenumbers.  The growth rate is normalized by the angular
velocity $\Omega$.  Three contour lines indicate $\gamma/\Omega = 0.7$
({\it solid curve}), 0.4 ({\it dashed curve}), and 0.2 ({\it
dot-dashed curve}).  In this figure, we assume the shear parameter $q
= 1.5$, the Brunt-V\"ais\"al\"a frequency $N = 0$, the ratio of the
angular velocity to the Alfv{\'e}n frequency $\Omega/\omega_A=30$, and
the field distribution $s = - 2$.}
\label{figure1}
\end{figure}

First we show the dependence of the growth rate of an unstable branch 
on the shear parameter $q$. 
The unstable growth rate is given by the imaginary part of $\omega$, i.e.,
$\gamma \equiv \Im (\omega)$.
In this subsection, the Brunt-V\"ais\"al\"a frequency is assumed to be
$N =0$.
Figure~\ref{figure1} is a contour plot of the growth rate $\gamma$ for
a strong differential rotation case with $q = 1.5$.
The growth rate is shown as a function of the 
azimuthal wavenumber $m$ and the ratio of poloidal wavenumbers 
$\beta \equiv k_r/k_z$.
In this figure, the azimuthal wavenumbers are treated as real values
instead of integer.
As is expected, the NMRI modes can be seen for this case and the
unstable growth rate is of the order of the angular velocity $\Omega$.
The maximum growth rate is $\gamma_{\rm{max}} \approx 0.75\Omega$ at
$\beta = 0$ and $m = 29$.
For a given $m$, the growth rate takes the maximum at $\beta=0$. 
The modes with a larger azimuthal wavenumber are stabilized 
with increasing $\beta$. 
When $k_r \ne 0$, the radial displacements of perturbations lead to
undulations of the magnetic pressure, which acts as a stabilizing
force for NMRI.
Only when $m$ is small there exists unstable modes with a larger
$\beta \gg 1$, although the growth rate is much smaller than $\Omega$.

\begin{figure}
\epsscale{.80}
\plotone{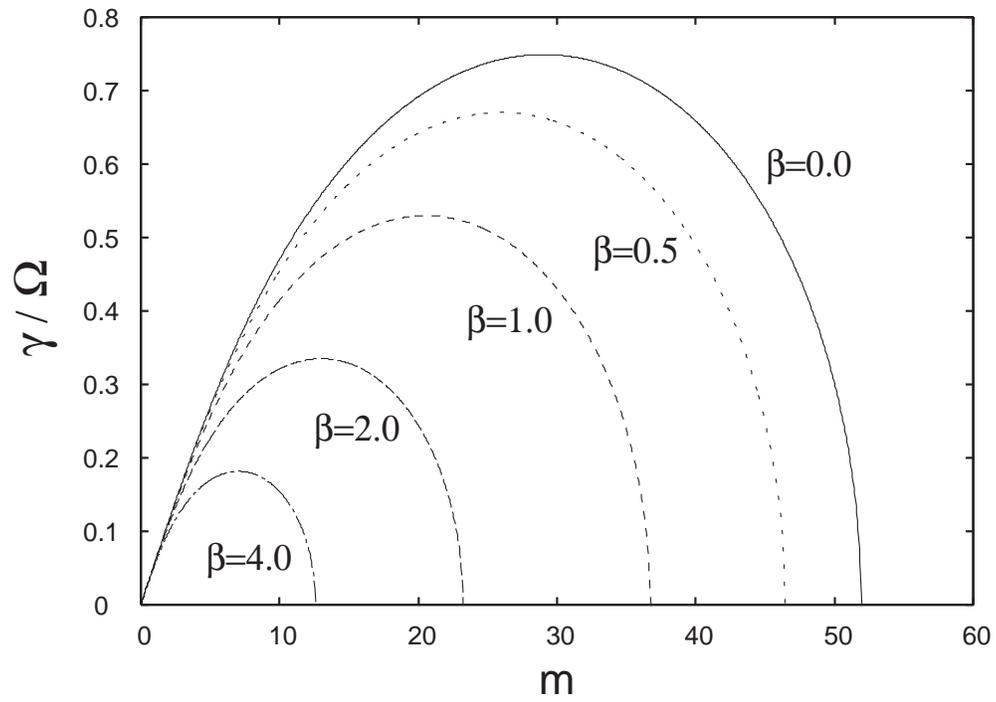} 
\caption{The unstable growth rate $\gamma/\Omega$ as a function of the
azimuthal wavenumber $m$.  Each lines are labeled by the ratio of the
poloidal wavenumbers $\beta$.  Model parameters are the same as those
in Figure 1.}
\label{figure2}
\end{figure}

Figure~\ref{figure2} shows the growth rate $\gamma$ as a function of 
the azimuthal wavenumber $m$ for $\beta = 0$, 0.5, 1, 2, and 4. 
Model parameters are the same as those in Figure~\ref{figure1}. 
The shape of the dispersion relation is quite similar to that of the 
axisymmetric MRI (Balbus \& Hawley 1991).
For a given $\beta$, unstable modes exist when the azimuthal
wavenumber $m$ is smaller than a critical value $m_{\rm crit}$ and 
the growth rate takes the maximum at $m_{\rm{max}}$. 
These characteristic wavenumbers and the maximum growth rate decreases
as $\beta$ increases.
When $\beta = 0$, the critical and the most unstable wavenumbers are
$m_{\rm{crit}} = 51$ and $m_{\rm{max}} = 29$, respectively.

It would be helpful to recall the axisymmetric MRI to understand the physical
meanings of the characteristic values $m_{\rm{max}}$, $m_{\rm{crit}}$,
and $\gamma_{\rm{max}}$ shown in Figures~\ref{figure1} and
\ref{figure2}.
The vertical wavenumber of the fastest growing mode $k_{\rm{max}}$
and the critical wavenumber $k_{\rm{crit}}$ for the axisymmetric MRI are
given by $k_{\rm{max}}  =  \sqrt{-q^2 + 4q}\ \Omega /2v_{\rm{A}}$ and
$k_{\rm{crit}}  =  \sqrt{2q}\ \Omega/v_{\rm{A}}$, and 
the maximum growth rate is $\gamma_{\rm{max}}  =  q\Omega/2$ (Balbus
\& Hawley 1991). 
When the vertical wavenumber $k$ is replaced by $m/r$, we can estimate
the azimuthal wavenumbers $m_{\rm{max}} = \sqrt{-q^2 + 4q}\
\Omega/2\omega_{\rm{A}} \approx 29$ and $m_{\rm{crit}} = \sqrt{2q}\
\Omega/\omega _{\rm{A}} \approx 51$ for $q =1.5$ and
$\Omega/\omega_{A} = 30$. 
The maximum growth rate is given by $\gamma_{\rm{max}} = q\Omega/2 =
0.75\Omega$ for $q = 1.5$. 
These values are exactly consistent with those derived from our dispersion
equation (\ref{eq22}).

\begin{figure}
\epsscale{.80}
\plotone{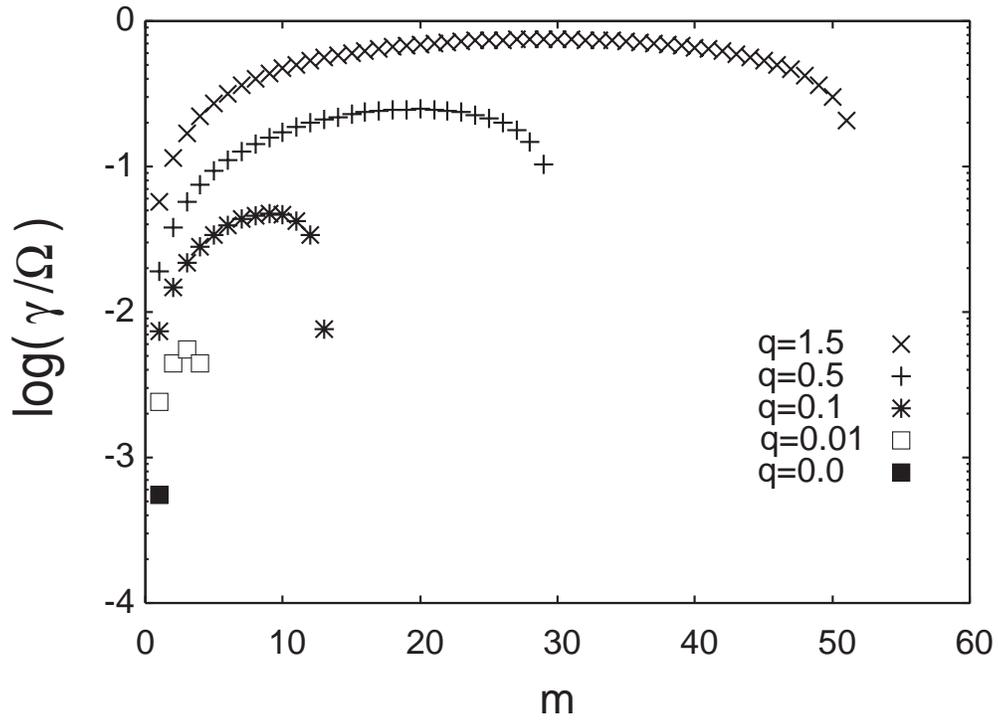}
\caption{The dependence of the growth rate on the shear parameter $q$.  The
growth rate is shown as a function of the azimuthal wavenumber $m$ for
the cases of $q$ = 1.5, 0.5, 0.1, 0.01, and 0. The poloidal wavenumber is 
assumed to be $\beta = 0$. The other parameters
are the same as in Figure 1.  The growth rate decreases as the shear
parameter decreases.  Only the $m=1$ mode can be unstable for the
rigid rotation case ($q = 0$).}
\label{figure3}
\end{figure}

Figure~\ref{figure3} shows the growth rate $\gamma$  
as a function of the azimuthal wavenumber $m$ for the cases with
different shear parameters $q = 0$, 0.01, 0.1, 0.5, and 1.5. 
The Brunt-V\"ais\"al\"a frequency and the poloidal wavenumber are
assumed to be $N=0$ and $\beta =0$.
As seen from this figure, the growth rate of NMRI decreases as the
shear parameter $q$ decreases.
When $q \sim 1$, the growth rate is comparable to the angular
velocity.
However, if $q$ is less than unity, the maximum growth rate is 
almost proportional to $q$.
The characteristic wavenumber $m_{\max}$ becomes smaller as the
shear is weaker.
When $q \lesssim 10^{-3}$, only $m=1$ mode is unstable and the the
growth rate is about $10^{-3} \Omega$.
As is described above, the maximum growth rate of axisymmetric MRI 
is proportional to $q$, and thus it reaches to zero at $q = 0$. 
However nonaxisymmetric $m=1$ mode is still unstable even for the
rigid rotation case ($q = 0$).
This is quite different from the axisymmetric case and a unique
feature of the NMRI.
 
\subsection{Kink-type (Tayler) instability}
The stability of the system with a toroidal magnetic field 
and a weak differential rotation is investigated by many authors 
(Tayler 1973; Acheson 1978; Schmitt \& Rosner 1983; Pitts \& Tayler 1986). 
They found that the kink-type instability ($m=1$) would be important
in rigidly rotating stellar interior. 
The instability criterion for the $m\neq 0$ mode depends 
on the radial distribution of the field, 
$s = -d \ln B_{\phi}/ d \ln r$ (Spruit 1999).  
The condition for the instability is given by 
\begin{equation}
 s <  - \frac{m^2}2 -1 \;.
\label{eq24}
\end{equation}
Obviously the kink mode ($m = 1$) is unstable when $s=-2$. 
By comparison, the instability criterion for the 
pinch mode ($m = 0$) is given by 
\begin{equation}
s < - \dsf{1}{2}\left(\frac{\Omega}{\omega_A}\right)^2 - 1 \;.
\label{eq24a}
\end{equation}
The pinch mode ($m = 0$) is usually stable in rigidly rotating
stellar interiors if $\Omega \gg \omega_A$. 

The growth rate of the kink-type instability is of the order of 
$\omega^2_A/\Omega$ (Spruit 1999), which is about $10^{-3} 
\Omega$ for the case of $\Omega/\omega_A = 30$. 
Thus the unstable mode in the limit of $q \rightarrow 0$ shown in
Figure~\ref{figure3} must be the kink-type instability. 
Notice that the NMRI and kink-type instability are 
on the same branch of the dispersion equation 
and the kink-type instability
corresponds to an asymptotic solution of the NMRI 
in the limit of the rigid rotation ($q \rightarrow 0$). 
The growth rate of the kink-type instability is much smaller than 
that of the NMRI in PNSs with a strong differential rotation, 
so that it would be anticipated that the NMRI would be dominant there. 

Hereafter, we describe the unstable modes with $m=1$ as "kink modes" 
in distinction from "NMRI modes" of $m \ge 2$. 
The marginal shear rate is estimated by using typical growth rates of
the NMRI modes ($\sim q\Omega$) and the kink modes ($\sim
\omega^2_A/\Omega$). From the comparison of these growth rates, we can
easily find that 
NMRI modes are dominant in the strong differential rotation cases 
$q \gg (\omega_A/\Omega)^2$, while the kink modes are dominant in the
weak differential rotation cases $q \ll (\omega_A/\Omega )^2$.

\subsection{Effects of the Brunt-V\"ais\"al\"a Frequency} 
\begin{figure}
\begin{center}
\begin{tabular}{c}
\scalebox{0.35}{\rotatebox{270}{\includegraphics{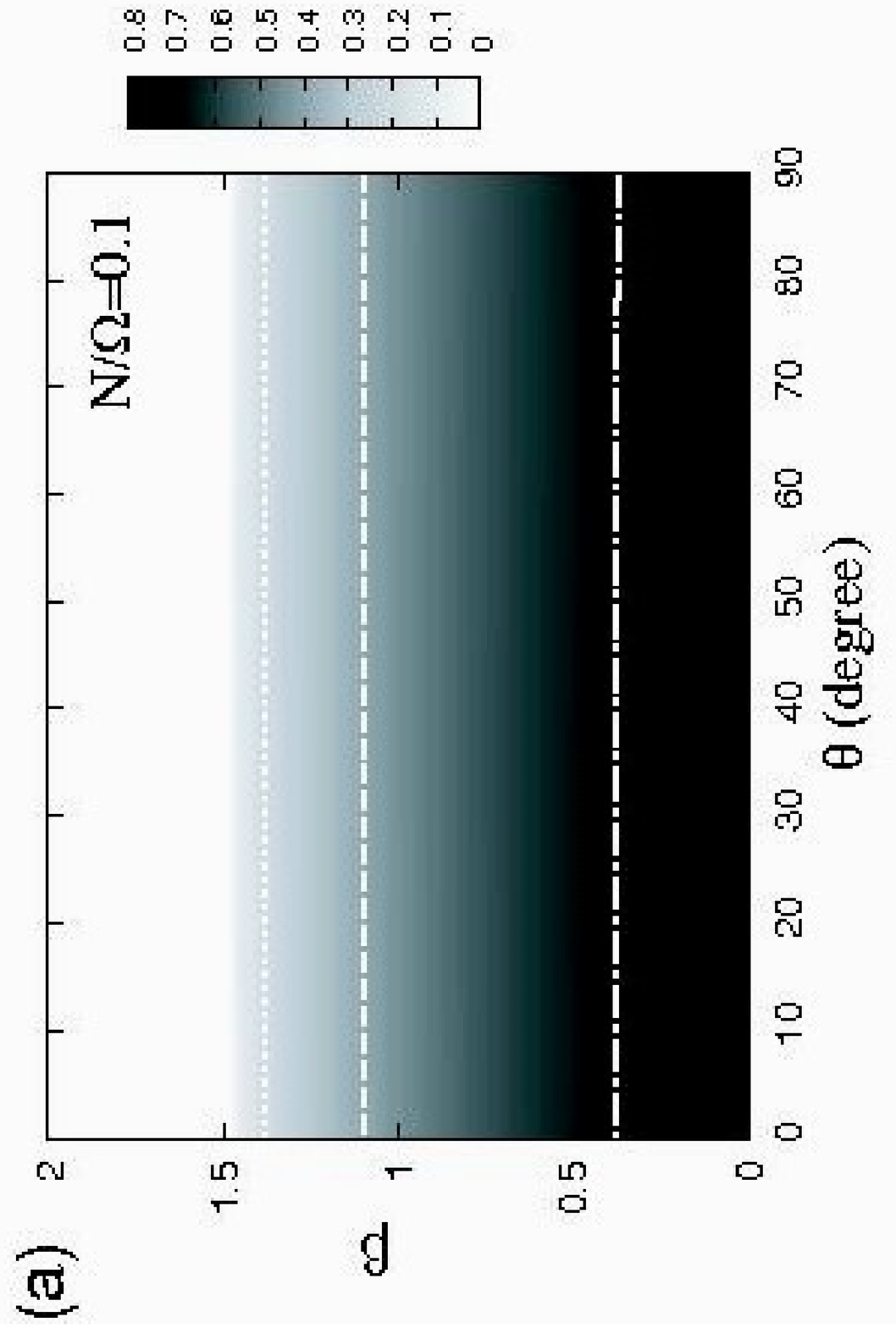}}}\\
\scalebox{0.35}{\rotatebox{270}{\includegraphics{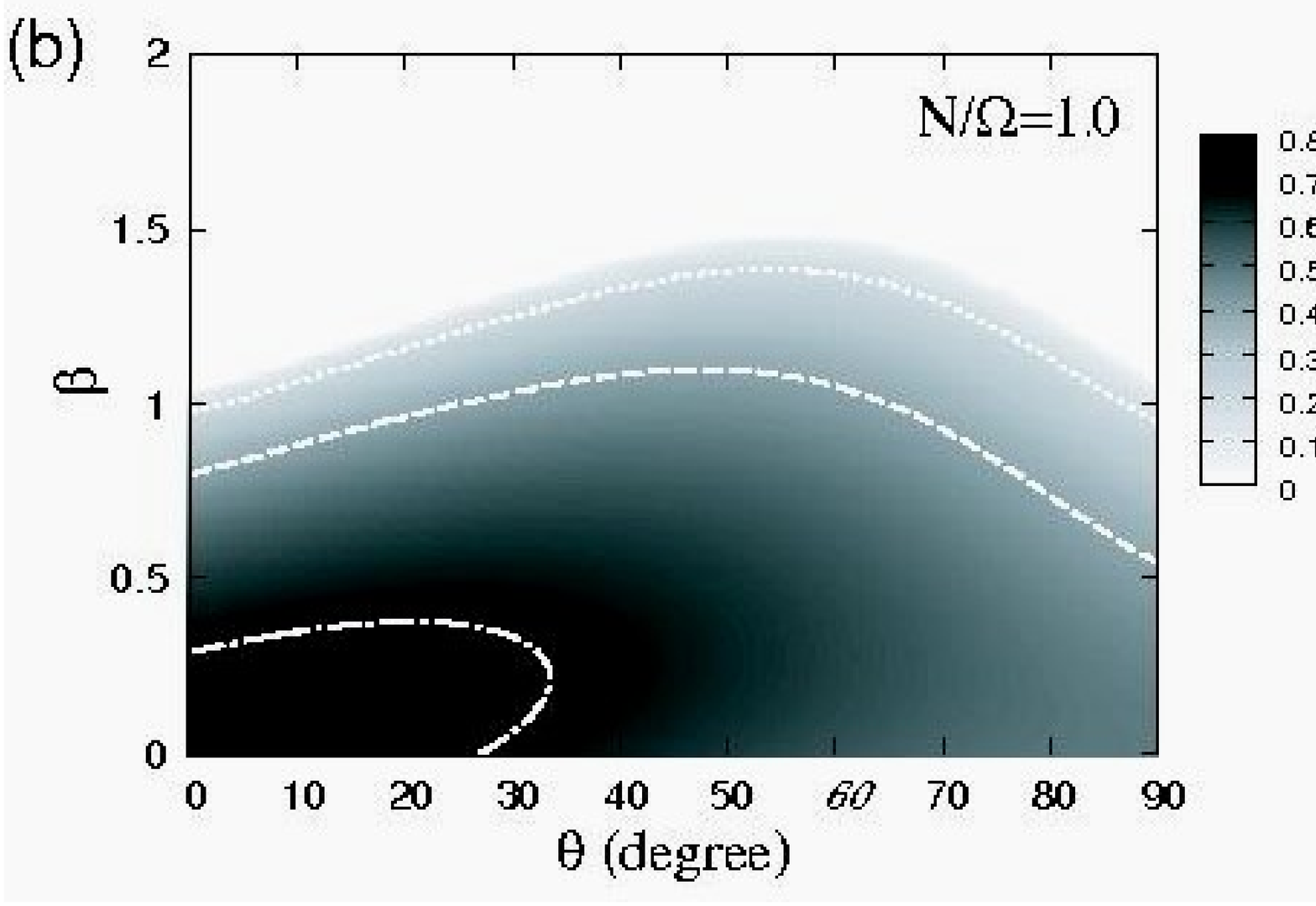}}}\\
\scalebox{0.35}{\rotatebox{270}{\includegraphics{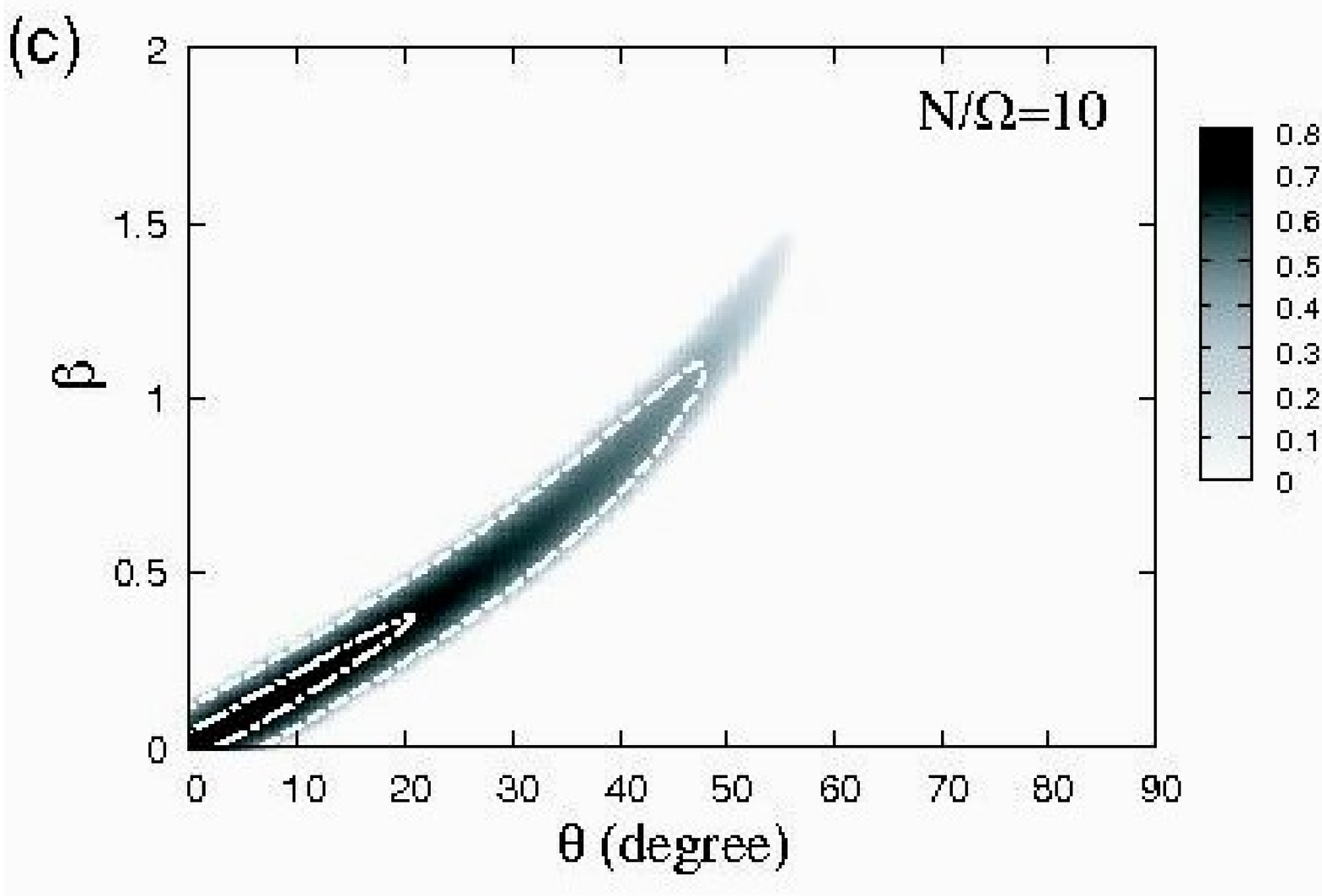}}}\\
\end{tabular}
\caption{The contour plot of the growth rate of the NMRI modes ($m = 29$) for
the case of (a) $N/\Omega = 0.1$, (b) $N/\Omega = 1.0$, and (c)
$N/\Omega = 10$.
The growth rate is shown on the $\theta$-$\beta$ diagram.
The polar angle $\theta$ is defined as $\theta \equiv \tan^{-1} (r/z)$
and measured with a unit of degree.  The contour lines in each figure
denote $\gamma/\Omega = 0.7$ ({\it dot-dashed curve}), 0.4 ({\it dashed
curve}), and 0.2 ({\it dotted curve}).  Model parameters are the same as those
in Figure 1 except for the Brunt-V\"ais\"al\"a frequency $N$.}
\label{figure4}
\end{center}
\end{figure}

In this subsection, we investigate the effects of the 
Brunt-V\"ais\"al\"a frequency $N$ on the NMRI and kink modes. 
The Brunt-V\"ais\"al\"a frequency $N$ is real at stably stratified 
regions.
The restoring force acts on displaced fluid elements along the
direction of the entropy and leptonic gradients 
$\mbox{\boldmath $N$}$, which is 
usually toward the center of the star.
Using a polar angle $\theta \equiv \tan^{-1}(r/z)$, the radial
and vertical components of the Brunt-V\"ais\"al\"a frequency are
written as
\begin{equation}
N_r=N\sin\theta ,\kern 15pt N_z=N\cos\theta \;. \label{eq25} 
\end{equation}
Figure~\ref{figure4} shows the effects of $N$ on the NMRI modes. 
The growth rate is shown by a gray contour on the $\theta$-$\beta$
plane for the case of $m=29$, which corresponds to the fastest
growing mode of the NMRI. 
Since the size of the Brunt-V\"ais\"al\"a frequency in PNSs is highly
uncertain (see \S~\ref{sec:application}), we investigate a wide range
of $N/\Omega$. From top to bottom, Figure~\ref{figure4} shows the
results for
$N/\Omega =0.1$, 1, and 10. 
The shear parameter is assumed to be $q = 1.5$.

The growth rate of the NMRI decreases with increasing $N$. 
The stabilizing effect by the Brunt-V\"ais\"al\"a oscillation depends
on the angle between the poloidal wavevector $\mbox{\boldmath $k$}$
and $\mbox{\boldmath $N$}$. 
When $N/\Omega =0.1$ (Fig.~\ref{figure4}a), the effect of $N$ is weak, 
so that the growth rate is independent of $\theta$. 
The mode with $\beta=0$ is always the fastest and the maximum growth
rate is $0.75\Omega$. 
When $N$ is comparable to $\Omega$ (Fig.~\ref{figure4}b), the growth
rate of the modes with $\mbox{\boldmath $k$} \perp \mbox{\boldmath
  $N$}$ is slightly suppressed due to the restoring force of $N$. 
On the other hand, unstable modes with the poloidal wavevector parallel to
$\mbox{\boldmath $N$}$ is unaffected by the Brunt-V\"ais\"al\"a
oscillation and still have a growth rate of the order of the angular
velocity $\Omega$. 
If the Brunt-V\"ais\"al\"a frequency is much larger than $\Omega$
(Fig.~\ref{figure4}c), the unstable modes can exist only when
$\mbox{\boldmath $k$} \parallel \mbox{\boldmath $N$}$.
Near the equatorial plane ($\theta \ga 60\arcdeg$), 
the growth of the NMRI ($m=29$) is completely suppressed.

\begin{figure}
\begin{center}
\begin{tabular}{c}
\scalebox{0.35}{\rotatebox{270}{\includegraphics{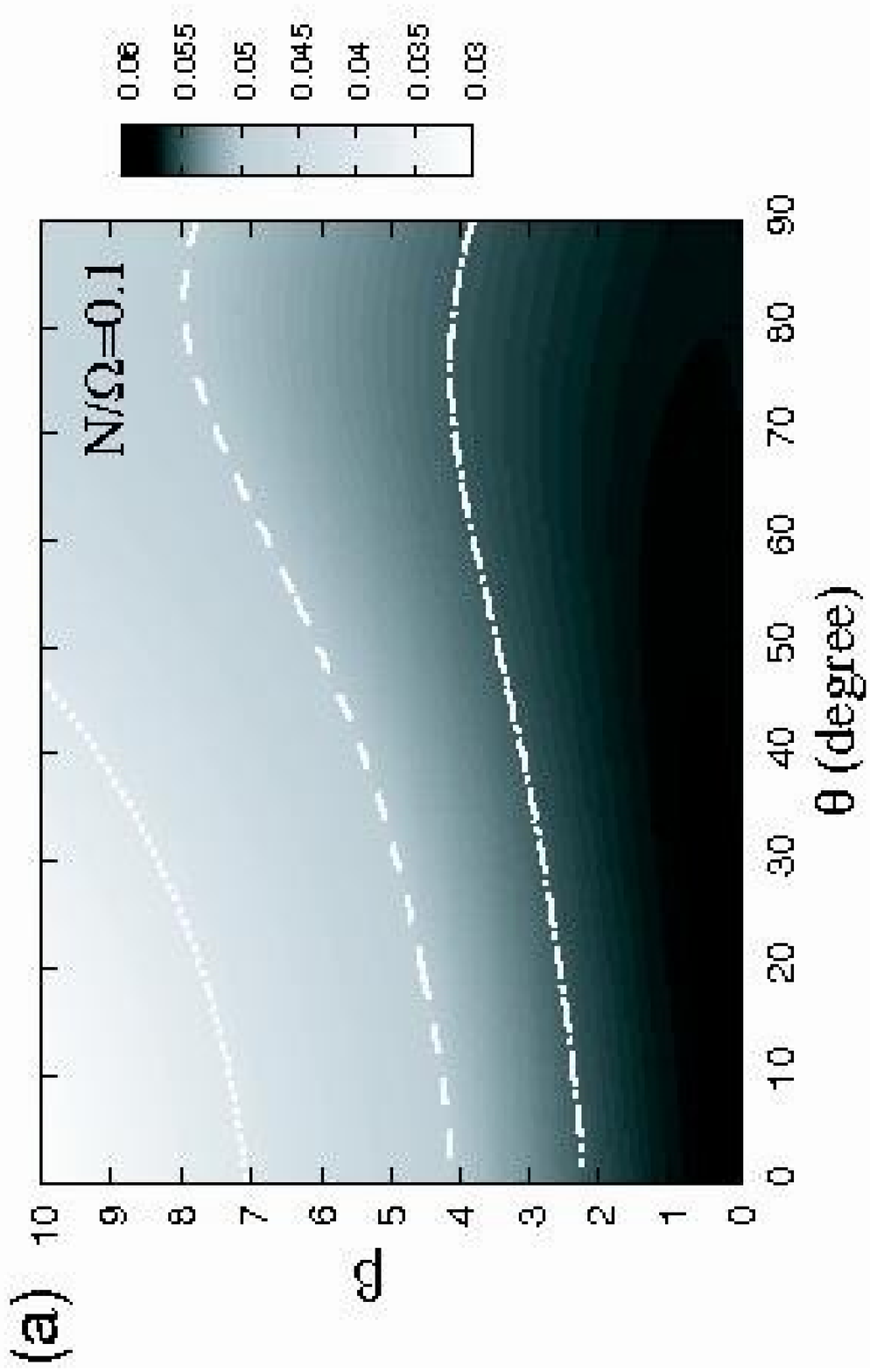}}} \\
\scalebox{0.35}{\rotatebox{270}{\includegraphics{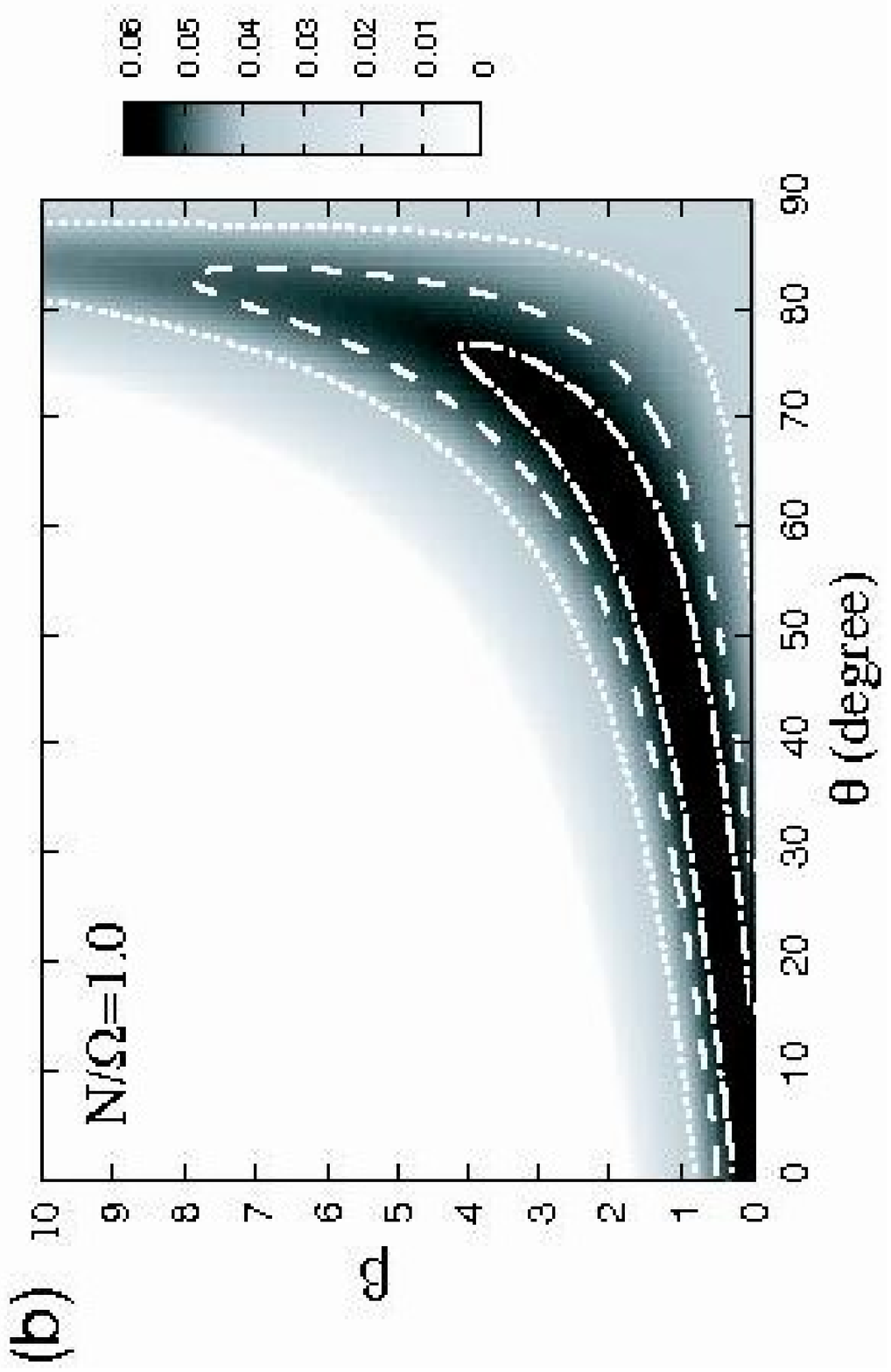}}} \\
\scalebox{0.35}{\rotatebox{270}{\includegraphics{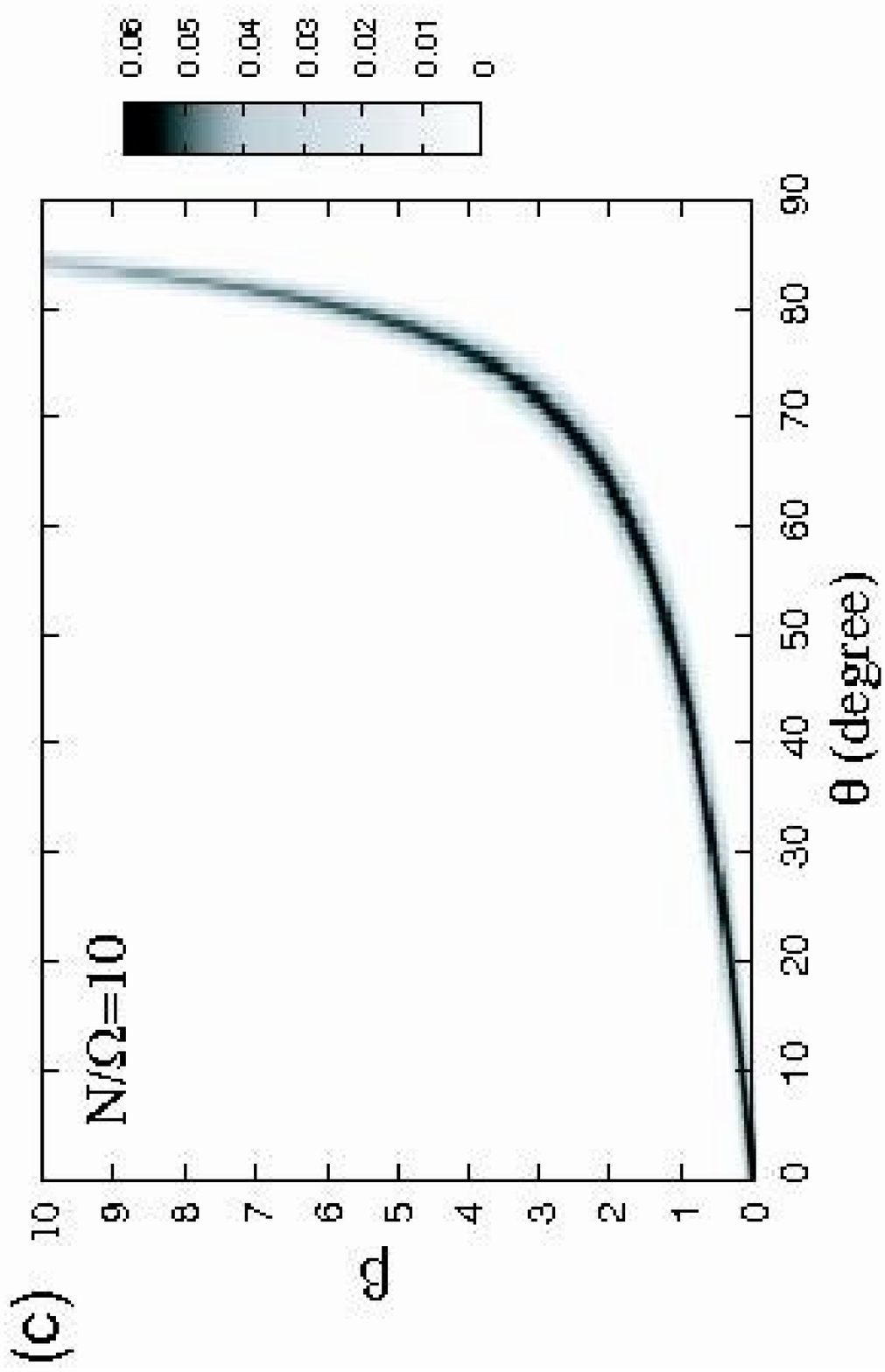}}} \\
\end{tabular}
\caption{The contour plot of the growth rate of the kink modes ($m = 1$) for
the case of (a) $N/\Omega = 0.1$, (b) $N/\Omega = 1.0$, and (c)
$N/\Omega = 10$.
The growth rate is shown on the $\theta$-$\beta$ diagram.
The polar angle $\theta$ is defined as $\theta \equiv \tan^{-1} (r/z)$
and measured with a unit of degree.  The contour lines in each figure
denote $\gamma/\Omega = 0.055$ ({\it dot-dashed curve}), 0.05 ({\it dashed
curve}), and 0.04 ({\it dotted curve}). Model parameters are the same as those
in Figure 1 except for the Brunt-V\"ais\"al\"a frequency $N$. }
\label{figure5}
\end{center}
\end{figure}

Next, we examine the effect of $N$ on the kink modes. 
Figure~\ref{figure5} shows the dimensionless growth rate
$\gamma/\Omega$ as a function of $\beta$ and $\theta$ for the case of
$m=1$.  
Notice that the growth rate of {kink modes} is much smaller than that
of {NMRI modes}.
When $N$ is comparable or larger than $\Omega $, 
almost all modes except that $\mbox{\boldmath $k$}$ is parallel to
$\mbox{\boldmath $N$}$ are suppressed, the same as in the case of
{NMRI modes}. 

The difference between the NMRI and kink modes
can be seen in the vicinity of the equatorial plane. 
The NMRI modes are suppressed around equatorial plane 
with increase of $N/\Omega$ (Fig.~\ref{figure4}c).
On the other hand, the kink modes can survive 
even near the equatorial plane (Fig.~\ref{figure5}c). 
The direction of the entropy and leptonic gradients is almost
horizontal when $\theta
\sim 90 \arcdeg$, so that only a mode with a wavevector $\mbox{\boldmath
  $k$} \parallel \mbox{\boldmath $N$}$, or $\beta = k_r/k_z \gg 1$,
can grow against the Brunt-V\"ais\"al\"a oscillation.
As seen in Figure~\ref{figure1}, the kink mode can exist in the limit 
of $\beta \gg 1$, while the NMRI is suppressed completely.
For the growth of the NMRI, 
the radial displacement of a perturbed fluid element is essential. 
However, the radial motion is stabilized near the equatorial plane and 
the NMRI cannot grow in such the situation. 
This would be the origin of the different behavior between two 
instabilities near the equatorial plane. 

\section{ANALYTICAL TREATMENT}
In this section, we discuss the stability criteria and the 
maximum growth rate using an analytical approach. 
Since the dispersion equation (\ref{eq22}) has a very complex form, 
it is difficult to treat for what it is. 
Therefore we simplify it under realistic approximations. 
Our simplified dispersion equation captures the features 
of the original equation precisely, 
so that our analytical stability criteria would be
useful for applications to the stellar interiors. 

The stability criteria for the cases of a weak differential rotation,
$q \ll (\omega_A/\Omega)^2$, and a rigid rotation, $q = 0$, have
already been investigated in terms of the interior of normal stars
(Fricke 1969; Tayler 1973; Acheson 1978; Schmitt \& Rosner 1983; Pitts \& Tayler
1986; Spruit 1999, 2002). 
Here we consider a strong differential rotation case, $q \gg
(\omega_A/\Omega)^2$, because it would be common in PNSs.
In that case, NMRI modes are dominant and then we can use the high
rotational frequency approximation assuming that the growth rate is
comparable to the angular velocity and much larger than the Alfv\'en
frequency, $\omega \sim \Omega \gg \omega_A$. 

\subsection{Stability Criteria}
The dispersion equation (\ref{eq22}) can be rewrite as 
\begin{equation}
\frac{k^2}{k_z^2} \left( \omega^2 - m^2 \omega^2_A \right)^2 
- \left[ - 2 (1 - s) \omega^2_A + 
\kappa^2 + N^2_e \right] \left( \omega^2 - m^2 \omega^2_A \right)
- 4 \omega^2_A 
\left( \omega + m \Omega \right)^2 = 0 \;.
\label{eq26}
\end{equation}
We consider a case of high rotational frequency, $\omega \sim
\Omega \gg \omega_A$, and a large azimuthal wavenumber, $m \gg 1$.
Then the Alfv\'en frequency in the second term can be negligible 
compared to the epicyclic and Brunt-V\"ais\"al\"a frequencies. 
We can also approximate $\omega + m \Omega \sim m \Omega$ in the third term. 
Using these approximations, we can obtain a simplified dispersion equation,
\begin{equation}
\frac{k^2}{k^2_z} \left( \omega^2 - m^2 \omega^2_A \right)^2  - \left(
\kappa^2 + N^2_e \right) \left( \omega^2 - m^2 \omega^2_A \right)
 - 4 m^2 \omega^2_A \Omega^2 = 0 \;.
\label{eq27}
\end{equation}
This is a quadratic equation of $\omega$ and the analytical
solution is 
\begin{equation}
\omega^2 =  m^2 \omega^2_A + \frac12 \frac{k^2_z}{k^2} \left\{
  \kappa^2 + N^2_e \pm \left[  \left(\kappa^2 + N^2_e \right)^2 
+ 16
  \frac{k^2}{k^2_z} m^2 \omega^2_A \Omega^2 \right]^{1/2} \right\}
\;. \label{eq29}
\end{equation}

The stability condition is given by $\omega^2 > 0$, that is,
\begin{equation}
\left(\frac{k_r}{k_z}\right)^2 A_2 + 
\left(\frac{k_r}{k_z}\right) A_1 + A_0 > 0 \;,\label{eq30}
\end{equation}
where
\begin{equation}
A_2 = m^2\omega^2_A + N^2_z \;,\ \ \ A_1 = -2N_rN_z \;,\ \ \ A_0 = N^2_r
-2q\Omega^2 + m^2\omega^2_A \;. \label{eq31}
\end{equation}
This inequality [eq.~(\ref{eq30})] is satisfied if and only if 
(i) $A_2 + A_0 > 0$ and (ii) $D = A^2_1 -4A_2A_0 < 0$ .
By using these two conditions, we can derive two criteria for stability;
\begin{equation}
\left( \frac{\omega_A}{\Omega} \right)^2 m^2  
-q + \frac{1}{2}\left( \frac{N}{\Omega} \right)^2 >  0 \;, \label{eq32}
\end{equation}
\begin{equation}
N^2 \omega^2_A m^2 - 2q\Omega^2 N^2_z > 0 \;. \label{eq33}
\end{equation}
These criteria are quadratic functions of the azimuthal 
wavenumber $m$. 
Thus, equations (\ref{eq32}) and (\ref{eq33}) are 
consistently satisfied for every azimuthal wavenumber 
in the case that zeroth order term of $m$ is positive. 
Therefore the general stability criteria can be derived as 
\begin{equation}
N^2 + {\frac{\partial\Omega^2}{\partial\ln r}} > 0 \;,\label{eq34}
\end{equation}
\begin{equation}
N^2_z \frac{\partial\Omega^2}{\partial \ln r} > 0 \;.  
\label{eq35}
\end{equation}

These criteria have forms similar to the Solberg-H{\o}iland criteria 
but with gradients of angular velocity replacing the gradients of 
specific angular momentum.
If the leptonic gradients are ignored, 
these criteria are exactly identical to those of the 
axisymmetric MRI (Balbus \& Hawley 1995).

\subsection{Maximum Growth Rate}
The maximum growth rate of NMRI can be derived from equation (\ref{eq29}). 
By differentiating partially with respect to $m$, 
the azimuthal wavenumber $m_{\rm{max}}$ that provides an
extremum is obtained as
\begin{equation}
m_{\rm{max}} = \frac{1}{4\omega_A\Omega}\frac{k_z}{k}\left[ 16\Omega^4 -
  \left( \kappa^2  + N^2_e \right)^2 \right]^{1/2} \;. \label{eq36}
\end{equation}
Substituting $m_{\rm{max}}$ into equation (\ref{eq29}), 
the maximum growth rate with respect to $m$ is given by
\begin{equation}
\gamma_{\rm{max}} = \frac{q\Omega}{2}\frac{k_z}{k}\left( 1 -
 \frac{1}{2q}\frac{N^2_e}{\Omega^2} \right) \;. \label{eq37} 
\end{equation}

We consider the characteristics of $\gamma_{\max}$ in two limiting
cases, $N=0$ and $N\neq 0$. 
When $N=0$, the unstable growth rate takes the maximum value,
\begin{equation}
\gamma_{\rm{max}} = \frac{q\Omega}{2} \;, \label{eq38}
\end{equation}
when $k_r / k_z = 0$ and 
\begin{equation} 
m_{\max}  =  \frac{\Omega }{2 \omega_A}\sqrt{-q^2 + 4q} 
\;. \label{eq39}
\end{equation}
These results are consistent with the characteristics of the original
dispersion equation (\ref{eq22}) discussed in \S~3.1. 

If $N\neq 0$, the ratio of the poloidal wavenumbers for the
fastest growing mode $\beta_{\rm{max}}$ can be obtained by
differentiating equation~(\ref{eq37}) with respect to
$\beta$.
Then $\beta_{\rm{max}}$ should satisfy the following equation,
\begin{equation}
\left( \frac{N_r}{N_z} - \beta_{\max} \right) 
\left( \beta_{\max}^2 + \frac{N_r}{N_z}\beta_{\max}
+ 2 \right) - 2q\beta_{\max}
\left( \frac{\Omega}{N_z} \right)^2 = 0  \;. \label{eq40} 
\end{equation} 
When $N \gg \Omega $, we can solve this equation approximately and the
solution is given by $\beta_{\rm{max}} \sim N_r/N_z$. 
This means that the poloidal wavevector of the fastest growing mode
is parallel to the entropy and leptonic gradients ($\mbox{\boldmath
  $k$} \parallel \mbox{\boldmath $N$}$).
This is also consistent with the results of the original dispersion
equation shown in Figure~\ref{figure4}. 

\section{APPLICATION TO PROTO-NEUTRON STARS}\label{sec:application}
In this section, 
we apply the simplified dispersion equation~(\ref{eq27}) 
to the interiors of PNSs and investigate the stability 
for the NMRI. 
In order to do that, we must know the amplitude of the physical 
parameters in the dispersion equation. 
Here we estimate physical quantities of PNSs 
based on the observations of pulsars and 
the numerical simulations of core-collapse supernovae. 

In the following we concentrate our discussions on the convectively 
stable regions in PNSs, which locates at the outer layer 
of PNSs (Janka \& M\"uller 1996). 
If this region is magnetohydrodynamically unstable, 
neutrino luminosities would be amplified by various nonlinear 
magnetic processes (see \S~\ref{sec:discussion}). 
The gain region behind the accretion shock is another site of 
MHD instabilities being important. 
Thompson et al. (2005) suggest that the 
viscous dissipation caused by MRI can enhance the heating in 
the gain region sufficiently enough to yield explosions. 
However, the gain region is convectively unstable. 
Convective motions may affect the growth of the MRI as is shown in 
convection-dominated accretion disks (Narayan et al. 2002). 
Although the linear and nonlinear interaction between the NMRI and 
convection would be a quite important issue, it is beyond the scope of 
this paper. 

\subsection{Physical Quantities in Proto-Neutron Stars}
PNSs are expected to rotate differentially, as opposite to old 
neutron stars. 
Even when the iron core of a supernova progenitor is almost rigidly
rotating (Heger et al. 2000; Heger et al. 2005), the
collapse can generate a significant amount of differential rotation. 
Numerical studies of the rotating core-collapse indicate that 
the outer layers of PNSs would be rotating at the angular velocity
$\Omega \simeq 100$ -- $1000 \ \rm{sec^{-1}}$ with a shear parameter 
$q \lesssim 1$ (Buras et al. 2003; Villain et al. 2004; Kotake et
al. 2004). 
This angular velocity is comparable to the observations of 
young isolated pulsars associated with supernova remnants (Marshall et
al. 1998). 
Thus, we adopt a rotational profile of 
$\Omega = 100\ \rm{sec^{-1}}$ and $q \simeq 1$ as typical values in PNSs. 

MHD simulations of the core-collapse suggest that 
toroidal magnetic fields would be much stronger than 
the poloidal component (Akiyama et al. 2003; Kotake et al. 2004; 
Yamada \& Sawai 2004; Takiwaki et al. 2004; Ardeljan et al. 2005). 
This is because the toroidal component is generated by wrapping of
poloidal fields by strong differential rotation during the
core-collapse phase. 
On the other hand, observations of young isolated pulsars 
indicate that poloidal magnetic fields of neutron stars are typically 
of the order of $10^{12-13}$ G. 
Thus we expect that the toroidal fields at 
the interiors of PNSs could be $10^{13}$ G. 
When we employ $B_{\phi} = 10^{13}$ G,
the Alfv\'en frequency is 
\begin{equation}
\omega_A \simeq 3.0 \left( \frac{B_{\phi}}{10^{13}\ \rm{G}} \right)
\left( \frac{r}{10^{6.5} \ \rm{cm}} \right)^{-1}
\left( \frac{\rho}{10^{11} \ \rm{g} \ \rm{cm}^{-3}} \right)^{-1/2} \
\rm{sec^{-1}} \;. 
\end{equation}


The size of the Brunt-V\"ais\"al\"a oscillation 
in the stably stratified layer of PNSs
is sensitive to microscopic physics such as 
the equation of state and neutrino opacity
(Buras et al. 2003; Kotake et al. 2004; Thompson et al. 2005; Dessart
et al. 2005). 
Therefore, we treat the Brunt-V\"ais\"al\"a frequency 
as a parameter. 
Because the Brunt-V\"ais\"al\"a frequency would be
larger than the Alfv\'en frequency in the
stellar interiors, $N^2 \gg \omega^2_A $, the models with
$N/\Omega = 0.1$ -- $10$ are examined here.
We use our simplified dispersion equation (\ref{eq27}) for the
stability analysis. 
This equation does not include the parameter $s$, so that we can
neglect the effects of field distribution in the following discussion.

\subsection{Stability of Proto-Neutron Stars}

\begin{figure}
\begin{center}
\begin{tabular}{c}
\scalebox{0.65}{\rotatebox{0}{\includegraphics{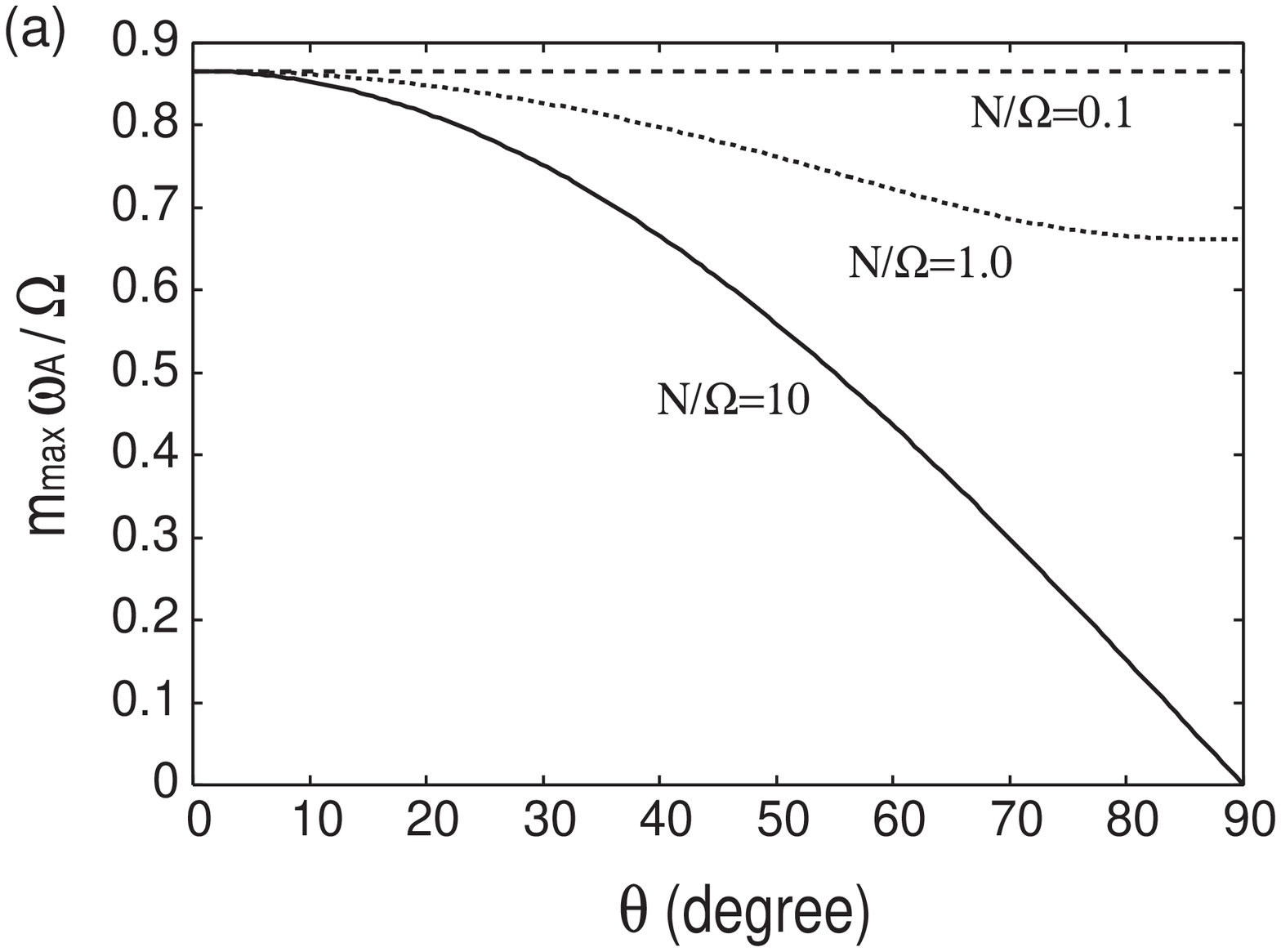}}} \\
\end{tabular}
\begin{tabular}{c}
\scalebox{0.65}{\rotatebox{0}{\includegraphics{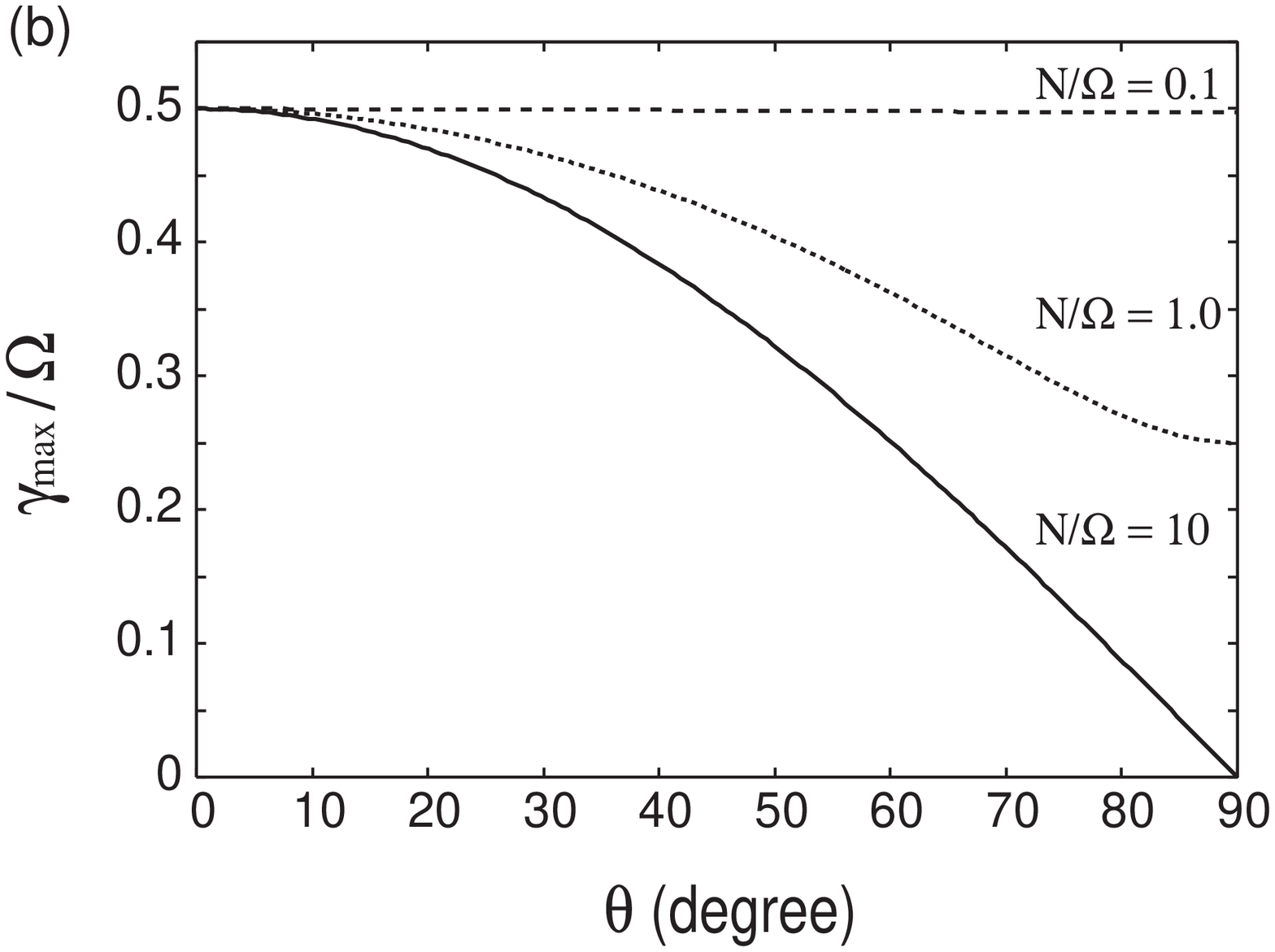}}} \\
\end{tabular}
\caption{(a) The azimuthal wavenumber and (b) the maximum growth rate of the
fastest growing mode as a function of the polar angle $\theta$ for the
cases with $N/\Omega = 0.1$, 1.0, and 10.
These are obtained from our simplified dispersion equation (\ref{eq27}).
We choose the model parameters as 
$q = 1.0$, $\Omega = 100 \ \rm{sec^{-1}}$, and $\omega_A = 3 \
\rm{\sec^{-1}}$. }
\label{figure6}
\end{center}
\end{figure}

The characteristics of the NMRI in stably stratified layers of PNSs 
are shown in Figure~\ref{figure6}. 
We assume the physical parameters $q = 1$, $\Omega = 100\
\rm{sec^{-1}}$, and $\omega_A = 3.0 \ \rm{sec^{-1}}$.
For the Brunt-V\"ais\"al\"a frequency, we examine three cases with 
$N/\Omega = 0.1$, 1, and 10.

Figure~\ref{figure6}a depicts the normalized azimuthal wavenumber 
of the fastest growing mode $m_{\rm{max}} \omega_A/\Omega$ obtained
from equations~(\ref{eq36}) and (\ref{eq40}).
The characteristic wavenumber $m_{\max}$ depends on the polar angle 
$\theta$ and the Brunt-V\"ais\"al\"a frequency $N$. 
Figure~\ref{figure6}b shows the dimensionless maximum growth rate 
$\gamma_{\rm{max}} / \Omega$ as a function of the polar angle $\theta$. 
The maximum growth rate $\gamma_{\rm{max}}$ is determined by 
$\beta_{\rm{max}}$ using equation~(\ref{eq37}). 
When the Brunt-V\"ais\"al\"a frequency is small $N/\Omega=0.1$, 
the azimuthal wavenumber $m_{\rm{max}}$ is independent of the polar 
angle $\theta$. 
The maximum growth rate $\gamma_{\rm{max}}$ 
is constant at a value given by equation (\ref{eq38}), which means 
that the NMRI is quite important in the entire regions. 
If the Brunt-V\"ais\"al\"a frequency $N$ is comparable to the angular 
velocity $N/\Omega = 1$, 
the maximum growth rate decreases near the equatorial plane. 
However the decrease of the growth rate is at most by a factor of 2 
at $\theta \sim 90 \arcdeg$. 
The azimuthal wavenumber is much larger than unity at any angle 
$\theta$ so that the NMRI is still dominant over the kink-type instability. 

Even when $N/\Omega$ is much larger than unity, all the regions are 
magnetohydrodynamically unstable. 
However the kink modes are dominant in the vicinity of the equator and 
thus the maximum growth rate decreases by many orders. 
Near the rotational axis, on the other hand, the azimuthal wavenumber 
$m_{\max}$ and the maximum growth rate $\gamma_{\rm{max}}$ are
unaffected by the Brunt-V\"ais\"al\"a frequency $N$. 
We define the critical angle $\theta_{\rm crit}$ at where the maximum 
growth rate becomes half of the value at $\theta = 0$.
The critical angle is about $\theta \sim 60 \arcdeg$ for the case of 
$N/\Omega =10$. 
We find that the critical angle is independent of the 
Brunt-V\"ais\"al\"a frequency if $N/\Omega \gtrsim 10$.

As a result, we can conclude that stably stratified
layers of PNSs are always unstable to MHD instabilities.
For the cases of $N/\Omega < 1$, the NMRI can grow at any angle
$\theta$. 
Even when $N/\Omega$ is larger than unity, the polar regions is
unstable to the NMRI. 
The growth rate is comparable to the angular velocity $\Omega$ and
much faster than that of the kink modes.
The kink modes can be dominant only in the vicinity of the equator 
when the Brunt-V\"ais\"al\"a frequency exceeds the angular velocity. 
The typical growth time of NMRI is about 100 msec.
This is much shorter than the neutrino cooling time of PNSs 
($\sim 10$ sec), and thus the nonlinear growth of NMRI may
affect the neutrino luminosities. 

We adopt $\Omega/\omega_A =30$ as a typical value in the discussions
above.
Recent numerical studies of rotating core-collapse suggests
the possibility of more rapidly rotating PNSs up to 
$\Omega \simeq 1000$ sec$^{-1}$ (Fryer \& Heger 2000; Dimmelmeier et
al. 2002; Thompson et al. 2005) and more strongly magnetized PNSs
with $B_{\phi} \simeq 10^{15}$ G as an origin of magneters
(Kotake et al. 2004; Yamada \& Sawai 2004;  Takiwaki et al. 2005;
Sawai et al. 2005).
Therefore, it would be interesting to consider the 
rapidly rotating case, $\Omega/\omega_A \gtrsim 100$, and 
strongly magnetized case, $\Omega/\omega_A \simeq 1$.

The growth rate of the NMRI is proportional to the angular velocity.
For rapidly rotating PNSs ($\Omega/\omega_A \gtrsim 100$), 
the NMRI can grow in shorter time scale than 10 msec.
The characteristic length scale in the azimuthal direction, $v_{\rm A}
/ \Omega$, shifts to the shorter wavelength.
In strongly magnetized PNSs ($\Omega/\omega_A \simeq 1$), on the other
hand, the NMRI is stabilized by the strong magnetic tension. 
The critical azimuthal wavenumber for the growth 
of NMRI is given by $\sqrt{2q}\ \Omega/\omega_A$, and thus the NMRI
modes would be suppressed completely 
if the field is very strong $\omega_{\rm A} \sim \Omega$.
Only the kink modes can survive in such a situation because they are not
affected by the magnetic tensions.

\section{DISCUSSION}\label{sec:discussion}
\subsection{Effects of Diffusion Processes} 

In the interior of PNSs, neutrino radiation plays an important 
role in the heat and lepton transports. 
The thermal and lepton diffusions caused by the neutrino radiation 
can reduce both the thermal and leptonic buoyant frequencies
significantly. 
The coefficients of the thermal and chemical 
diffusivities in PNSs are given by  
\begin{equation} 
\kappa_{\rm{T}} \simeq 10^9 \ 
Y^{-1}_{1/3} l^2_{30} \frac{T_4}{\mu^{6}_{20}} 
\ {\rm cm}^2 \ \sec^{-1}
\;, \ \ \ 
\nu_{\rm{chem}} \simeq 10^9 \ l^2_{30} \frac{T_4}{\mu^{6}_{20}} 
\ \ {\rm cm}^2 \ \sec^{-1} \;, \label{eq45} 
\end{equation} 
where $Y_{1/3}$ is the electron fraction in units of $1/3$, 
$l_{30}$ is a length scale in units of 30 km, 
$T_4$ is the temperature in units of 4 MeV, 
and 
$\mu_{20}$ is the electron chemical potential
normalized by 20 MeV (Socrates et al. 2005).

These diffusion processes must be considered if the wavelength of
fluctuations is shorter than the critical length scale that
can be estimated by equating the diffusion time and the period of the
buoyant oscillations. 
The critical length scale of the thermal diffusion is given by 
$\lambda_{\rm{T}} \sim \sqrt{\kappa_{\rm{T}}/ N_{\rm{T}}} \simeq 10^3$
cm, where we assume $N_{T} \simeq 1000$ sec$^{-1}$.
The critical length of the chemical diffusion 
may be a few times larger than that of the thermal one, 
because the leptonic buoyant frequency is smaller than $N_{\rm T}$
(Thompson \& Murray 2000). 
Thus, our analysis is valid for the wavelength of the
perturbations longer than the diffusion length $\lambda_{\rm T}$.

However, the diffusive processes could affect the growth of the NMRI
for the shorter wavelength $\lambda < \lambda_{\rm{T}}$.
The primary effect of the thermal and chemical diffusions is to reduce
the restoring force caused
by the stable stratification (Acheson 1978; Spruit 1999). 
Therefore, the maximum growth rate of NMRI increases and approaches
to  the case of $N/\Omega \rightarrow 0$ when $\lambda \ll \lambda_{\rm T}$. 
Thus, 
the growth of these shorter wavelength modes may be dominant 
in the stably stratified layers of PNSs.

\subsection{NMRI Modes vs. Kink Modes}
\begin{figure}
\begin{center}
\begin{tabular}{c}
\scalebox{0.35}{\rotatebox{270}{\includegraphics{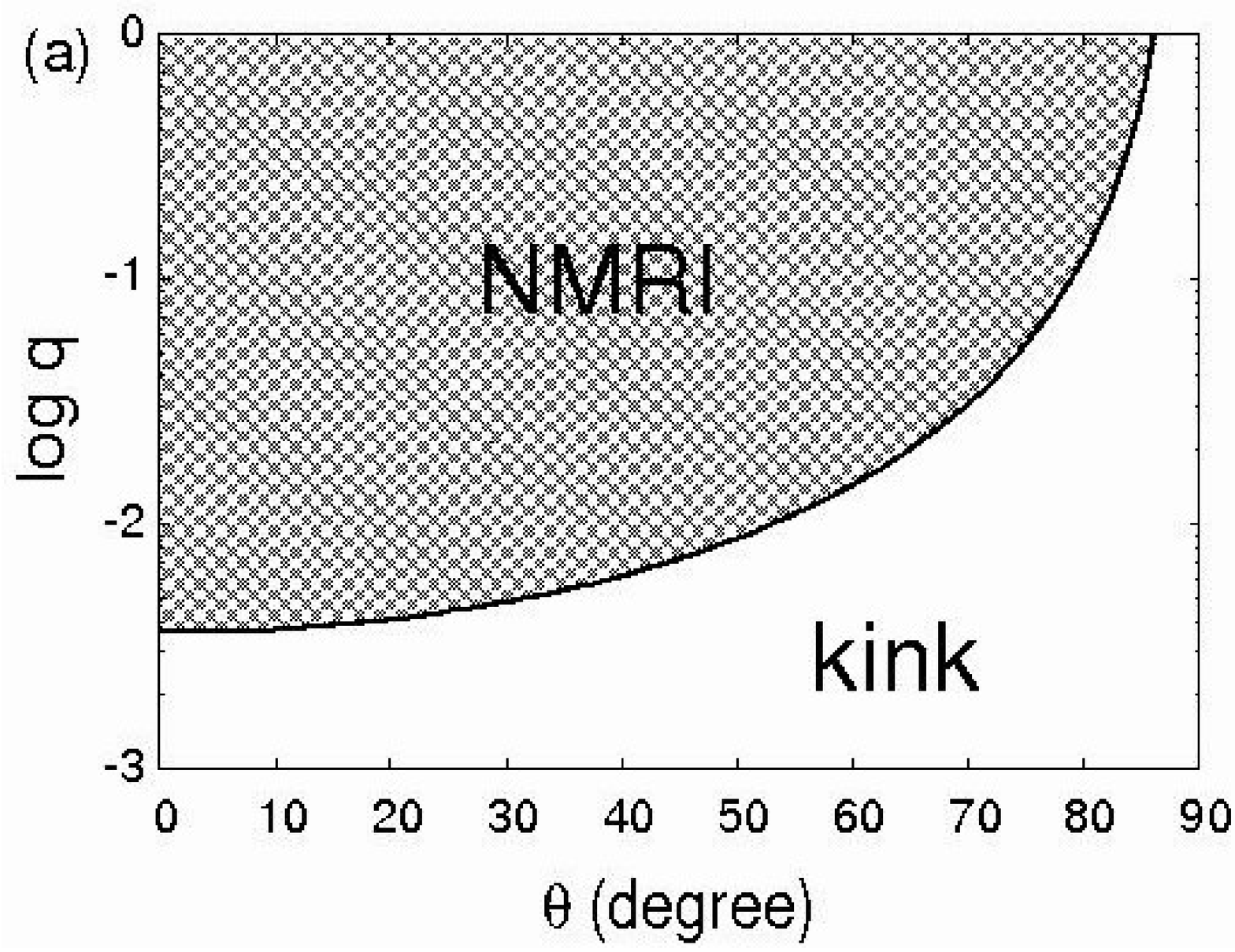}}} \\
\end{tabular}
\begin{tabular}{c}
\scalebox{0.35}{\rotatebox{270}{\includegraphics{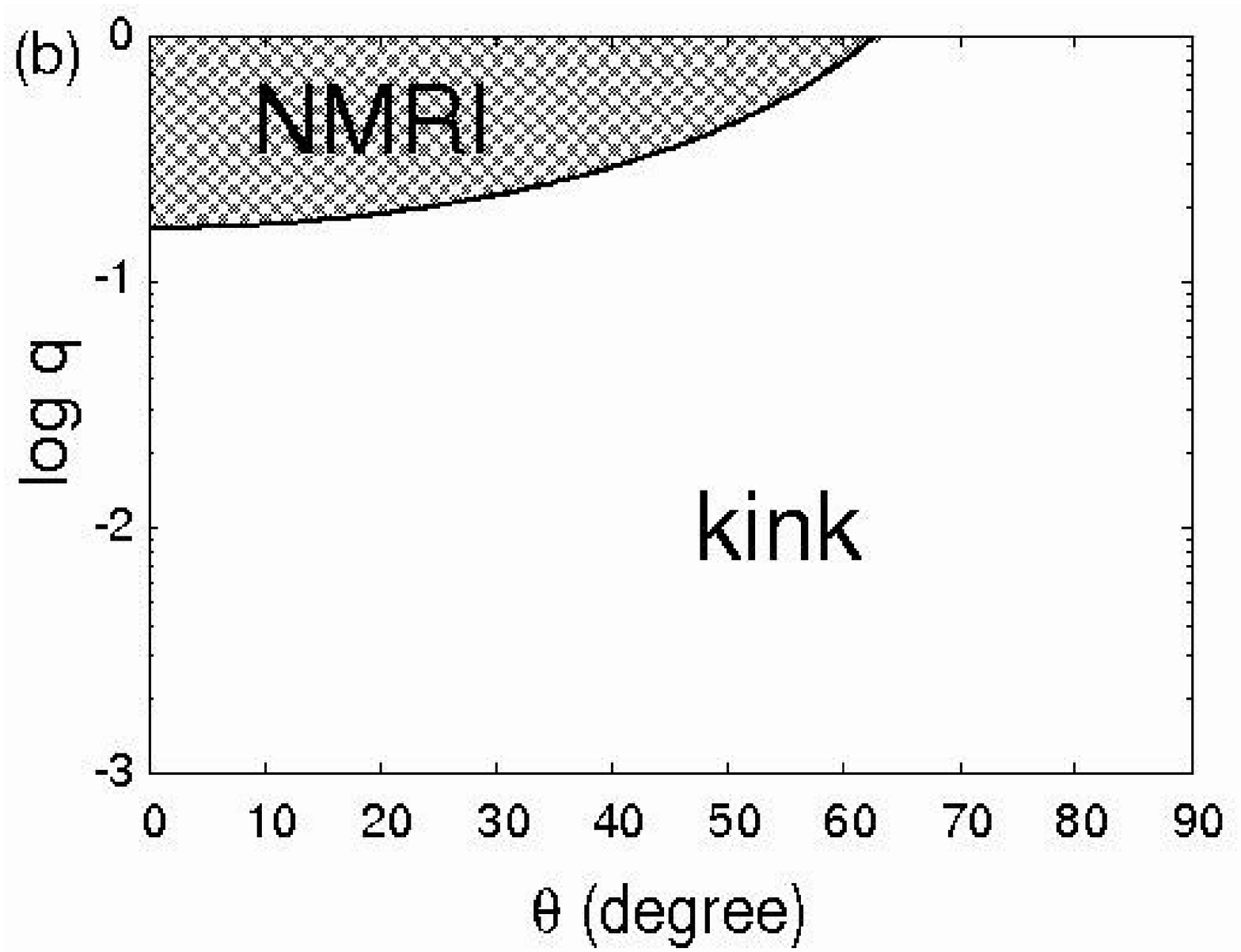}}} \\
\end{tabular}
\caption{(a) The NMRI and kink dominant regimes in the diagram of the shear
parameter $q$ and the polar angle $\theta$ for the case of the PNSs.
The border of the two regions is where the maximum azimuthal
wavenumber is $m_{\max} = 2$ in equation (\ref{eq36}).
Model parameters for this case are $N = 10 \Omega$ and $\Omega = 30
\omega_A$. 
(b) The same plot for the case of the solar radiative zone.
Model parameters are $N = 10^3 \Omega$ and $\Omega = 5 \omega_A$. }
\label{figure7}
\end{center}
\end{figure}

The conditions that the NMRI modes dominate over the kink modes can be 
derived from our simplified dispersion equation~(\ref{eq27}). 
The NMRI modes grow faster if the azimuthal wavenumber $m_{\max}$ is 
more than 2. 
The shaded (white) area in Figure~\ref{figure7}a represents the NMRI 
(kink) dominant region in the $q$-$\theta$ diagram. 
The boundary of these two areas is given by where 
$m_{\rm{max}} = 2$ is fulfilled in equation~(\ref{eq36}). 
In this figure, we assume $N/\Omega=10$ and $\Omega/\omega_A=30$ 
which are the typical values in PNSs. 
The NMRI modes are dominant at most of the regions if the shear 
parameter $q$ is relatively large ($q \gtrsim 10^{-2}$). 
The kink modes become important only when the shear parameter is 
extremely smaller than unity. 
The boundary curve is little changed by the size of the 
Brunt-V\"ais\"al\"a frequency. 

For comparison, a similar picture for the solar radiative zone is 
shown by Figure~\ref{figure7}b. 
In the solar radiative zone, 
the Brunt-V\"ais\"al\"a frequency 
is much larger than the angular velocity $N/\Omega \simeq 10^3$ (Spruit 2002). 
Here we assume the Alfv\'en frequency 
$\Omega/\omega_A = 5$ as the typical value in the solar radiative zone. 
The parameter region where the kink modes are dominant extends to the 
larger $q$ regime because of the smaller ratio of $\Omega/\omega_A$. 
Only when the shear rate is about unity, 
the NMRI modes can be important for this case. 
According to the helioseismology, the solar radiative zone exhibits 
a nearly rigid rotation ($q \ll 1$). 
Thus, the kink modes are dominant and Tayler-Spruit dynamo would be 
amplifying the magnetic fields (Spruit 2002). 
\subsection{Nonlinear Growth of the NMRI}
The NMRI can grow in the stably stratified layer of PNSs 
regardless of the size of the Brunt-V\"ais\"al\"a frequency. 
The nonlinear growth of the NMRI could initiate and 
sustain MHD turbulence and amplify magnetic fields. 
As is well known in the contexts of accretion disks, 
MHD turbulence leads to the angular momentum transport.
Thus the rotational configurations within PNSs evolve 
toward a rigid rotation. If the PNSs will rotate rigidly, 
the neutrino sphere become oblate and the neutrino luminosies 
could be enhanced in the polar region (Kotake et al. 2003). 
Thus, the polar region would be heated by neutrinos preferentially.

Turbulent viscosity generated by the NMRI can convert the 
rotational energies into the thermal energies, 
so that the neutrino luminosities from the PNSs 
may be increased (Thompson et al. 2005). 
Ramirez-Ruiz \& Socrates (2005) suggest that the heating of 
materials through magnetic reconnection may also lead to the 
enhancement of neutrino luminosity. 
We would like to stress that the magnetoconvection 
and magnetic Rayleigh-Taylor instability of magnetic flux tube 
would assist the mixing of materials in the layers 
where is heretofore thought to be convectively stable 
(Parker 1974, 1975, 1979). 

Although these effects are absolutely a matter of speculation, 
it is sure that these effects are not considered 
in the present scenarios of core-collapse supernovae. 
Therefore, we anticipate that these multiple magnetic effects can 
enhance the neutrino luminosity sufficiently to yield explosions 
in models that would otherwise fail. 
Because these magnetic phenomena are essentially three-dimensional, 
we must perform three-dimensional MHD simulations to reveal those 
effects quantitatively. 
This is our future work. 

\section{SUMMARY}
In this paper, we perform the local linear analysis of 
the stably stratified layers in stellar objects taking account of
toroidal magnetic fields and differential rotation.
We derive the local dispersion equation and apply it to the interior
of PNSs. 
Our findings are summarized as follows.

1. 
NMRI modes are dominant when the rate of differential rotation $q$ is
large and the growth rate is comparable to the angular velocity.
In the limit of rigid rotation ($q \to 0$),
only $m = 1$ mode is unstable, which corresponds to the kink
instability. 
The growth rate of the kink mode is given by
$\omega^2_A/\Omega$ and typically much smaller than that of the NMRI.

2.
The growth of NMRI is suppressed by the Brunt-V\"ais\"al\"a
oscillation.
However, because the stabilizing force is along the direction of the
entropy and leptonic gradients $\mbox{\boldmath $N$}$, 
the unstable modes can exist if the wavevector is parallel to
$\mbox{\boldmath $N$}$. 
Thus the kink modes is dominant over the NMRI modes in the 
vicinity of the equator when $N/\Omega$ is large.

3. 
Analytical formula of the stability criteria for NMRI and 
the maximum growth rate are derived from our simplified dispersion
equation.
The stability criteria are completely identical to those for the
axisymmetric MRI if the leptonic gradient is ignored.
The maximum growth is also agreed with that of the axisymmetric MRI for
the case of $N=0$. 

4. 
NMRI can grow in the interiors of differentially rotating PNSs
in spite of the size of the Brunt-V\"ais\"al\"a frequency. 
The suppression by the Brunt-V\"ais\"al\"a oscillation can be seen
only near the equatorial plane where the kink mode is dominating.
The typical growth time of NMRI is about 100 m sec, which is much
shorter than the neutrino cooling time of PNSs. 


\acknowledgments
We thank Shoichi Yamada and Kazunari Shibata 
for helpful discussions. We also thank the anonymous referee for
constructive comments. 
TS is supported by the Grant-in-Aid (16740111, 17039005) from the
Ministry of Education, Culture, Sports, Science, and Technology of
Japan. 


\clearpage

\clearpage

\clearpage

\clearpage
\clearpage

\clearpage

\clearpage
\end{document}